\documentclass{IEEEtran}
\usepackage{cite}
\usepackage{amsmath,amssymb,amsfonts}
\usepackage{algorithmic}
\usepackage{graphicx}
\usepackage{booktabs} % IEEE 推荐使用 booktabs 进行表格美化
\usepackage{array}
\usepackage{comment}
\usepackage{textcomp}
\usepackage{xcolor}
\usepackage{stfloats} % 允许双栏图片放置在底部
\usepackage{caption}    % 控制 caption 对齐方式
\usepackage{makecell}
\usepackage{afterpage}  % 允许在当前页底部放置 figure*
\usepackage{url}

\def\BibTeX{{\rm B\kern-.05em{\sc i\kern-.025em b}\kern-.08em
    T\kern-.1667em\lower.7ex\hbox{E}\kern-.125emX}}
\begin{document}
%\title{Modeling of Coupling Matrix of Dipole Antenna Array Using Physics-Aware Convolutional Long Short-Term Memory}
%\title{Enhanced Mutual Coupling Modeling in Dipole Antenna Arrays Using PC-LSTM: A Physics-Aware Machine Learning Approach}
%\title{Enhanced Mutual Coupling Modeling in Dipole Antenna Arrays Using a Novel Physics-Aware Attention-Based Machine Learning Approach}
%\title{Novel Physics-Aware Attention-Based Machine Learning Approach for Mutual Coupling Modeling in Dipole Antenna Arrays}
\title{Novel Physics-Aware Attention-Based Machine Learning Approach for Mutual Coupling Modeling}
%\title{{\color{red}PC-LSTM: A Physics-Aware Attention-Based Machine Learning Approach for Mutual Coupling Modeling in Dipole Antenna Arrays}}
\author{Can Wang,  Wei Liu, \IEEEmembership{Senior Member, IEEE}, Hanzhi Ma, \IEEEmembership{Member, IEEE}, \\Xiaonan Jiang \IEEEmembership{Member, IEEE},  Erping Li, \IEEEmembership{Fellow, IEEE}, and Steven Gao, \IEEEmembership{Fellow, IEEE} 
\thanks{This work has been submitted to the IEEE for possible publication. Copyright may be transferred without notice, after which this version may no longer be accessible. The work is supported by the RGC Grant GRF 14210623, AoE/E-101/23-N, CUHK Startup Fund under Grant 4937125, Grant 4937126, Grant 8601743, Grant 4055185, and the National Natural Science Foundation of China under Grant No. 62431012.}
\thanks{Can Wang, Xiaonan Jiang, and Steven Gao are with the Department of Electronic Engineering, The Chinese University of Hong Kong, Shatin, Hong 
Kong (e-mail: 1155206738@link.cuhk.edu.hk;  xnjiang@link.cuhk.edu.hk; scgao@ee.cuhk.edu.hk). }
\thanks{Wei Liu is with the Department of Electrical and Electronic Engineering, Hong Kong Polytechnic University, Kowloon, Hong Kong (e-mail: wliu.eee@gmail.com).}
\thanks{Hanzhi Ma, and Erping Li are with the College of Information Science and Electronic Engineering, Zhejiang University, Hangzhou, China (e-mail: mahanzhi@zju.edu.cn; liep@zju.edu.cn).}
}

\maketitle
\begin{abstract}
This article presents a physics-aware convolutional long short-term memory~(PC-LSTM) network for efficient and accurate extraction of mutual impedance matrices in dipole antenna arrays.  
By reinterpreting the Green's function through a physics-aware neural network and embedding it into an adaptive loss function, the proposed machine learning-based approach achieves enhanced physical interpretability in mutual coupling modeling. Also, an attention mechanism is carefully designed to calibrates complex-valued features by fusing the real and imaginary parts of the Green’s function matrix. These fused representations are then processed by a convolutional long short-term memory network, and the impedance matrix of the linear antenna array can be finally derived. Validation against five benchmarks underscores the efficacy of the proposed approach, demonstrating accurate impedance extraction with up to a $7\times$ speedup compared to CST Microwave Studio, making it a fast alternative to full-wave simulations for mutual coupling characterization.
\end{abstract}

\begin{IEEEkeywords}
Mutual coupling, dipole antenna array, impedance matrix, self-attention mechanism, physics-aware convolutional long short-term memory
\end{IEEEkeywords}

\section{Introduction}
\label{sec:introduction}
\IEEEPARstart{D}{ipole} antenna arrays are prevalently implemented in various wireless systems due to their simple structure, low cost, and favorable radiation characteristics~\cite{wen2021compact,haupt2010antenna}. They serve as fundamental building blocks in numerous practical applications, including  the massive multiple-input multiple-output (MIMO) systems\cite{b6}, radar~\cite{b13,b14}, and satellite communication systems~\cite{b11,b12}. As these systems evolve, the demand for large-scale and high-density antenna arrays continues to grow, making the accurate modeling of such arrays increasingly critical~\cite{haupt2010antenna}.
In particular, when antenna elements are placed in close proximity, mutual coupling becomes a critical issue due to strong electromagnetic interactions~\cite{b15}. Mutual coupling distorts the radiation patterns and array manifold, leading to performance degradation in beamforming~\cite{b21,PreviousTAP}, direction-of-arrival (DoA) estimation~\cite{10738292}, and interference suppression~\cite{10858644}. Consequently, developing accurate and computationally efficient methods for mutual coupling modeling is crucial to ensure the reliability and performance of modern high-density antenna array systems~\cite{9411296} \cite{lu2024generalized} \cite{zhu2023multimode}.

%To this end, Various computational electromagnetic~(CEM) methods have been proposed to model mutual coupling in antenna arrays~\cite{b30, b31, b32,9714820}.  For instance, Debdeep~\textit{et al.} explored the finite-difference time-domain~(FDTD) algorithm for the design of the mutually coupled antennas~\cite{9714820}. Rubio~\textit{et al.} combine the three-dimensional finite element method~(FEM) and spherical mode expansion to calculate the overall generalized scattering matrix of the finite antenna array~\cite{1406246}.Gupta~\textit{et al.} proposed a modified steering vector to analyze coupling effects in Applebaum-type arrays~\cite{b25}.  Zhang~\textit{et al.} introduced a beam pattern synthesis method incorporating mutual coupling compensation and magnitude constraints, allowing for flexible pattern designs with efficient use of degrees of freedom~\cite{b27}. Yuan~\textit{et al.} developed a hybrid method the combine the method of moments (MoM) and geometrical theory of diffraction (GTD) to investigate mutual coupling in an adaptive array of monopole antennas mounted on a rectangular conducting plate~\cite{704828}. The challenges of these methods suffer from high computational complexity due to repeated iterations, requiring extensive computation time and memory for large-scale arrays, and exhibit limited effectiveness for non-uniform arrays \cite{b28, b36}. 

To this end, various computational electromagnetic  methods have been proposed to model mutual coupling in antenna arrays~\cite{9714820},\cite{1406246},\cite{b25},\cite{704828}. For instance, Debdeep~\textit{et al.} applied the finite-difference time-domain method to the design of mutually coupled antennas~\cite{9714820}. Rubio~\textit{et al.} combined the three-dimensional finite-element method with spherical mode expansion to compute the generalized scattering matrix of finite antenna arrays~\cite{1406246}. Gupta~\textit{et al.} proposed a modified steering vector to analyze coupling effects in Applebaum-type arrays~\cite{b25}. Yuan~\textit{et al.} developed a hybrid approach combining the method of moments (MoM) with the geometrical theory of diffraction to investigate coupling in adaptive monopole arrays mounted on a conducting plate~\cite{704828}.
These methods have demonstrated high accuracy and physical fidelity. However, their applications to large-scale arrays are constrained by computational complexity, memory consumption, and limited scalability~\cite{b28, b36}. 
With the rapid advancement of artificial intelligence, neural networks have shown to be able to approximate complex nonlinear functions with high accuracy and fast inference speed~\cite{b37}. Artificial neural networks (ANNs) have been applied to a range of mutual coupling tasks, including modeling and compensation~\cite{b36}, coupling prediction~\cite{b44}, and decoupling~\cite{b45, b46}. Many of these methods use physical or geometric parameters as inputs to directly learn the mapping to desired output quantities~\cite{b36, b42}. While effective in certain settings, their performance may be affected by the growing complexity of the mapping and increasing data requirements in large-scale array scenarios~\cite{b38}. Furthermore, purely data-driven models may lack physical interpretability, which can limit their generalization and robustness~\cite{b43} . 

%To address these challenges, physics-informed neural networks (PINNs), as introduced by Raissi~\cite{b43}, integrate governing physical equations into the training process. This hybrid learning paradigm enhances data efficiency, improves generalization, and offers greater interpretability in both forward and inverse modeling of antenna array systems~\cite{b47, b48}.

In this paper, we propose a novel physics-aware convolutional long short-term memory (PC-LSTM) framework for mutual coupling modeling in dipole antenna arrays. This physics-aware approach autonomously models the mutual coupling between array elements with minimal reliance on labeled data by embedding physical constraints into the learning process, thus enhancing learning efficiency. Specifically, a physics-aware neural network (PANN) is used to predict the Green's function, incorporating physical constraints into the loss function to improve accuracy and accelerate computation. The method then synthesizes the port impedance matrix of a two-port dipole antenna array with arbitrary spacing. An attention-based scheme is introduced to address the imbalance between the real and imaginary parts. 
Finally, the port impedance matrix for an arbitrary large-scale linear antenna array can be derived through the proposed PC-LSTM method.
the framework is applied to derive the port impedance matrix of a large-scale linear antenna array. Numerical examples are provided to demonstrate the effectiveness of the proposed approach.
The contributions of this work are listed as follows: 
\begin{itemize}
  \item [1)]
A novel physics-aware attention-based machine learning approach, namely PC-LSTM, is proposed for scalable and highly accurate port impedance modeling across half-wavelength dipole arrays of varying sizes, applicable to both uniformly and non-uniformly spaced configurations. Compared to the full-wave electromagnetic simulation software CST Microwave Studio \cite{cst2025}, the proposed method achieves over a 7$\times$ speedup, significantly reducing computational cost.
  \item [2)]
We reinterpret the Green’s function using a physics-aware neural network (PANN) and embed it into the loss function, thereby enhancing both physical interpretability and accuracy of mutual coupling modeling. In addition, we introduce an adaptive loss function to balance the real and imaginary components, which enables improved accuracy and faster convergence.
  \item [3)]
An attention mechanism is proposed to calibrate the complex-valued features of electromagnetic signals by adaptively fusing the real and imaginary components of the Green’s function matrix. The fused representations are subsequently processed by a convolutional LSTM (ConvLSTM) network, enabling accurate estimation of port impedance in two-element subarrays and large-scale arrays. This design enhances the expressiveness of feature representations and contributes to improved learning stability.
\end{itemize}

The remainder of this article is organized as follows. Section II presents the proposed method, which is verified by simulation results in Section III, and Section IV concludes this work.

\section{Methodology}
In this section, we first introduce the MoM for modeling mutual impedance. Then, we use a PANN to approximate the Green’s Function. Next, a self-attention mechanism is applied to calibrate complex features. We also design a special convolution kernel to capture features of impedance matrices. Finally, an LSTM network is employed to model impedance matrices of different lengths.
%Mutual coupling in antenna arrays, particularly under dense element spacing, is often caused by multiple incident electromagnetic waves interacting across elements. To model this effect with high fidelity, full-wave solvers such as the MoM are commonly employed. MoM converts integral formulations into solvable linear systems, yielding accurate solutions for induced surface currents. Herein, we integrate MoM as a physics-guided component to improve the accuracy and interpretability of the proposed machine learning framework. Hence, we briefly introduce the MoM process in the following.
\subsection{The Method of Moments}
To model mutual coupling effect with high fidelity, full-wave solvers such as the MoM are commonly employed. In particular, an incident electric field on a perfect electric conductor object or other conductive object induces surface currents, which subsequently generate scattered fields. The total electric field on the object's surface is obtained by~\cite{6236032}:
\begin{equation}
\mathbf{E}(\mathbf{r}) = \mathbf{E}_{\text{sc}}(\mathbf{r}) + \mathbf{E}_{\textbf{i}}(\mathbf{r})
\label{eq1}
\end{equation}
where \(\mathbf{E}\) is the total electric field, \(\mathbf{E}_{\textbf{i}}\) is the incident electric field, \(\mathbf{r}\) is the location of the observation point, and the scattered field \(\mathbf{E}_{\text{sc}}\) can be expressed in terms of the current via the Green's function \(G(\mathbf{r}, \mathbf{r}')\) as:
\begin{equation}
\mathbf{E}_{\text{sc}} = -j\omega\mathbf{A}-\nabla\phi
\label{eq2}
\end{equation}
where \(\mathbf{A}\) is the vector potential generated by the current density \(\mathbf{J}\), and \(\phi\) is the scalar potential associated with the charge density \(\rho\). Substituting \(\mathbf{A}\) and \(\phi\) into (\ref{eq2}) yields:
\begin{equation}
\mathbf{E}_{\text{sc}} = -j\omega\mu\int_{\Omega '} \Big[G(\mathbf{r}, \mathbf{r}')\mathbf{J}(\mathbf{r})'+\frac{1}{k^2}\nabla G(\mathbf{r},\mathbf{r}')\nabla' \cdot \mathbf{J}(\mathbf{r}')\Big]\,d\mathbf{r}'
\label{eq5}
\end{equation}
where \(k^2=\omega^2\mu\epsilon\), \(\mathbf{r}'\) is the position of the charge, \(\mu\) is the magnetic permeability, \(\epsilon\) is the electric permittivity, \(\omega=2\pi f\) is the angular frequency, \(f\) is the frequency, and the operator \(\nabla'\) represents that it only performs on source functions. To enable the computational solution of (\ref{eq5}), MoM is used to approximate its solution with a finite-dimensional subspace \cite{b30}. Also, a set of basis functions is selected to approximate the unknown current density \(\mathbf{J}(\mathbf{r}')\), denoted as \(\mathbf{J}(\mathbf{r}')=\sum_{i=1}^N I_n \psi_n(\mathbf{r}')\), where \(N\) is the number of subdivisions, \(I_n\) is an undetermined coefficient, and \(\psi_n(\mathbf{r}')\) represents the basis function used to approximate \(\mathbf{J}(\mathbf{r}')\). The piecewise linear rooftop basis function~\cite{b49, b50}, widely employed for 2D problems, is utilized in this paper due to its simplicity and effectiveness. By substituting this basis function into (\ref{eq5}), we can get:
\begin{align}
\mathbf{E}_{\text{sc}}(\mathbf{r}) &= -Z_0 \sum_{n=1}^N I_n jk\int_{\Omega '} \Big[G(\mathbf{r},\mathbf{r}')\psi_n(\mathbf{r}') \notag \\ 
&\quad + \frac{1}{k^2} \nabla G(\mathbf{r}, \mathbf{r}') \nabla' \cdot \psi_n(\mathbf{r}')\Big] d\mathbf{r}'
\label{eq8}
\end{align}
where \(Z_0\) is the characteristic impedance (\(Z_0=\mu c = \sqrt{\mu_0/\epsilon_0}\)), and \(c\) is the light speed (\(c=\frac{1}{\sqrt{\mu_0\epsilon_0}}\)), \(k\) is the wave number (\(k=\frac{\omega}{c}\)). By combining boundary conditions and Galerkin's method \cite{b30}, (\ref{eq1}) can be rewritten as:
\begin{align}
&\left\langle \psi_m (\mathbf{r}), \mathbf{E}_i (\mathbf{r}) \right\rangle = \int_{\Omega} \psi_m(\mathbf{r})\cdot\mathbf{E}_i(\mathbf{r})d \mathbf{r} \notag \\
&= Z_0 \sum_{n=1}^N I_n jk\int_{\Omega}\int_{\Omega '} \Big[G(\mathbf{r},\mathbf{r}')\psi_m(\mathbf{r})\psi_n(\mathbf{r}') \notag \\ 
&\quad - \frac{1}{k^2} G(\mathbf{r}, \mathbf{r}')\nabla\psi_m(\mathbf{r})  \nabla' \cdot \psi_n(\mathbf{r}')\Big] d\mathbf{r}'d\mathbf{r}
\label{eq9}
\end{align}

Then, the voltage \(V_m\) is defined as \(V_m = \int_{\Omega}\psi_m \cdot \mathbf{E}_i(\mathbf{r})d\mathbf{r}, m = 1,2,...,N\), and the corresponding impedance \(Z_{mn}\) is defined as:
\begin{align}
Z_{mn} & = jkZ_0 \int_{\Omega}\int_{\Omega '} \Big[G(\mathbf{r},\mathbf{r}')\psi_m(\mathbf{r})\psi_n(\mathbf{r}') \notag \\ 
&\quad - \frac{1}{k^2} G(\mathbf{r}, \mathbf{r}')\nabla\psi_m(\mathbf{r})  \nabla' \cdot \psi_n(\mathbf{r}')\Big] d\mathbf{r}'d\mathbf{r}
\label{eq11}
\end{align}
Based on \(V_m\) and (\ref{eq11}), (\ref{eq9}) can be rewritten in matrix form:
\begin{align}
\mathbf{V} &= \mathbf{Z}\mathbf{I}
\label{eq13}
\end{align}
where \(\mathbf{V}\) is the incident electric field projected onto the basis (can be seen as incident voltage), \(\mathbf{Z}\) is the impedance matrix representation of the Green’s function integral operator acting onto the basis, and \(\mathbf{I}\) is the unknown current projected on to the basis.

%The key to solving  (\ref{eq13}) lies in the computation of the impedance matrix, which is fundamentally based on the basis functions, the Green's function, and numerical integration. 
In (\ref{eq13}), as the Green's function characterizes the system's response to a unit source, it plays a vital role in evaluating the mutual interactions between basis functions. In free space, the scalar Green's function \(G(\mathbf{r}, \mathbf{r}')\) is given by:
\begin{equation}
G(\mathbf{r}, \mathbf{r}') = \frac{e^{-jk|\mathbf{r}-\mathbf{r}'|}}{4\pi|\mathbf{r}-\mathbf{r}'|}
\label{eq14}
\end{equation}

To improve the generalization capacity of the neural network, the frequency-dependent term is normalized to mitigate scale variations and enhance training stability, facilitating the extraction of underlying patterns across a wide frequency range. Herein, the frequency term is explicitly factored out in the following formulation:
\begin{align}
    G(\mathbf{r}, \mathbf{r}') &= \frac{e^{-jk|\mathbf{r}-\mathbf{r}'|}}{4\pi|\mathbf{r}-\mathbf{r}'|} = \frac{\text{exp}\left\{-j\frac{\omega}{c}|(m-n)\hat{\mathbf{a}}_r|\Delta l \right\}}{4\pi|(m-n)\hat{\mathbf{a}}_r|\Delta l} \notag\\
    & = \frac{\text{exp}\left\{-j\frac{\omega}{c}|\Delta_{mn}|\frac{\lambda}{2N}\right\}}{4\pi|\Delta_{mn}|\frac{\lambda}{2N}} = \frac{\text{exp}\left\{-j\frac{\pi}{N}|\Delta_{mn}|\right\}}{|\Delta_{mn}\frac{2\pi c}{fN}|} \notag\\
    & = k(f)\frac{\text{exp}\left\{-j\frac{\pi}{N}|\Delta_{mn}|\right\}}{|\Delta_{mn}|}, \forall m,n \in \{1,2,3, \dots, W\}
\label{gnf}
\end{align}
where \(\hat{\mathbf{a}}_r\) is the direction vector, \(\Delta l = l/N\) is the length of each unit, \(l=\lambda/2\) is the length of the dipole antenna, \(N\) is the number of subdivision units, \(\Delta_{mn}=|m-n|,(\forall m,n \in \{1,2,3, \dots, W\})\) represents the spacing, and \(k(f)=\frac{fN}{2\pi c}\) is the function of the frequency. (\ref{gnf}) enables a more structured representation of the problem and improves the network’s ability to learn underlying physical relationships. 
\begin{figure*}[!t] % * 让图片跨两列，b 让其尽量放在底部
\centering
\makebox[\textwidth]{ % 让图片整体居中
    \includegraphics[width=0.9\textwidth]{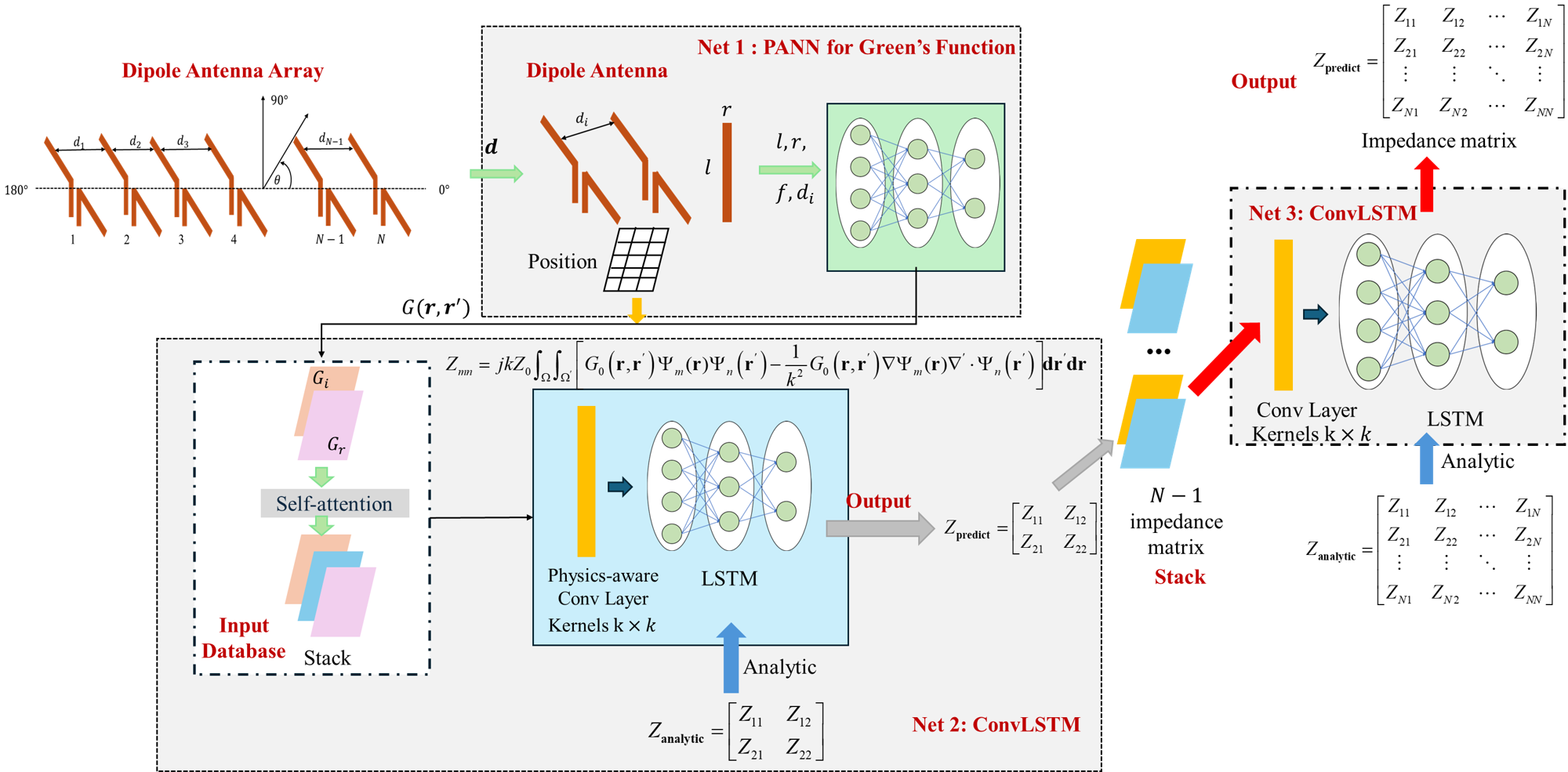} % 让图片接近两列宽度
}
%\captionsetup{justification=centering} % 让 caption 居中
\caption{Illustration of the proposed PC-LSTM architecture.} % 关键：手动添加 \centering
\label{fig2}
\end{figure*}
Due to the high computational complexity, limited scalability, and difficulties in addressing complex and multi-scale problems inherent to MoM, neural networks are increasingly adopted for electromagnetic modeling owing to their fast inference speed and strong generalization capability \cite{b51}. To effectively leverage the two methods' advantages for mutual coupling analysis in array systems, the core idea of our proposed method is illustrated in Fig. \ref{fig2}. For simplicity, dipole antennas and dipole antenna arrays are taken as examples. Firstly, the width \(r\), length \(l\), and operating frequency \(f\) are fed into PANN to generate the real and imaginary parts of the Green's function at each partitioned position. These outputs are fused into a new matrix through the self-attention mechanism. Next, together with the element spacing, the fused features and element spacing are first input to a physics-aware ConvLSTM to predict the impedance matrix of a two-element dipole array. This output is then processed by a second ConvLSTM to synthesize the impedance matrix of large-scale arrays in a cascaded manner.

\begin{figure}[!t]
\centerline{\includegraphics[width=\columnwidth]{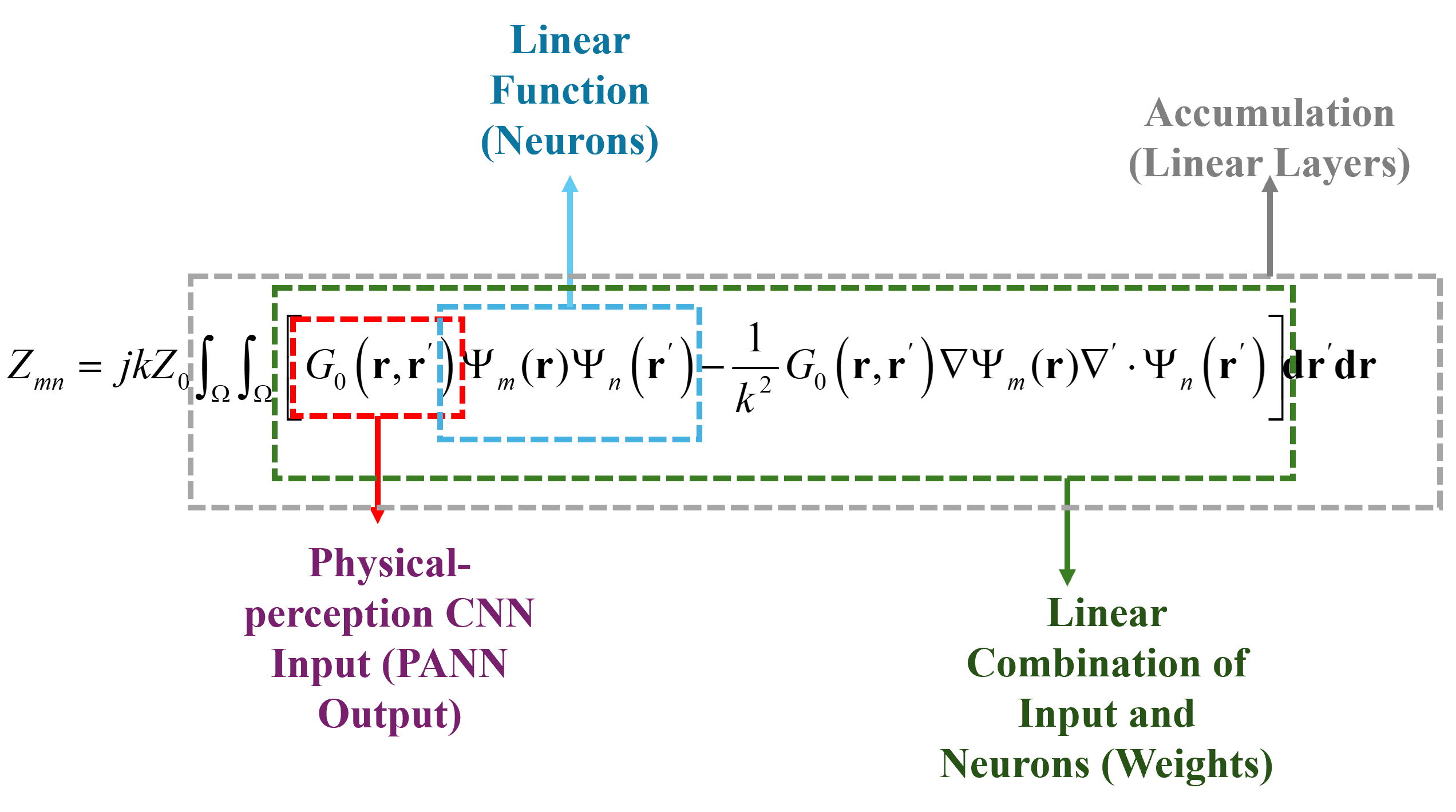}}
\caption{The correspondence between the MoM formulation and the neural network architecture.}
\label{fig3}
\end{figure}

%In our proposed method, the key mathematical {\color{red}equation (\ref{eq11})} is mapped to a neural network structure, as illustrated in Fig. \ref{fig3}. 

%This mathematical correspondence enables the proposed framework to be systematically divided into the following five parts.

\subsection{Physics-Aware Neural Network for Green's Function}
Physics-aware neural networks have demonstrated impressive improvements in applied electromagnetics, including beamforming~\cite{PreviousTAP}, uncertainty quantification~\cite{10701557}, metasurface design~\cite{su2025multi}, etc. In our scenario, as shown in Fig. \ref{fig3}, the impedance matrix in (\ref{eq11}) can be mapped to a neural network structure. Specifically, PANN generates the Green's function (red dashed box) and serves as input to the cascaded ConvLSTM, the activation function approximates the basis function (blue dashed box), and the combination coefficients (green dashed box) correspond to the network weights. In MoM, the integral over the computational domain is approximated by a discrete summation over small elements (gray dashed box), which aligns with the layer-wise accumulation in the neural network. In the following, PANN is introduced in detail.

By incorporating the analytical Green's function expressions from (\ref{gnf}), the loss function of PANN can be constructed as follows:
\begin{align}
\text{MSE}_{\text{PANN}} & = \frac{1}{N_s}\sum_{i=1}^{N_s} (y_{\text{pre}, i} - G(r_i, r_i'))^2 \notag \\
& = \frac{1}{N_s}||\mathbf{y}_{\text{pre}} - \mathbf{G}(\mathbf{r}, \mathbf{r}')||_2^2
\label{eq17}
\end{align}
where \(\mathbf{y}_{\text{pre}, i}\) represents the predicted output of PANN, \(N_s=N^2\) refers to the number of samples, \(N\) is the number of subdivisions, and the Green's function matrix \(\mathbf{G}(\mathbf{r}, \mathbf{r}')\) is directly from the analytical expressions in (\ref{gnf}), rather than being obtained from labeled training data. (\ref{eq17}) enables PANN to operate in an unsupervised learning framework, eliminating the need for labeled data and promoting better generalization by enforcing inherent physical constraints \cite{b53}.

Due to the significant order-of-magnitude discrepancy between the real and imaginary components, an adaptive loss function that dynamically adjusts the weighting to ensure a balanced optimization is proposed, enhancing the model's ability to learn complex-valued relationships effectively. The adaptive loss function is described as follows:
\begin{equation}
    L_{\text{total}}=\omega_rL_r+\omega_iL_i
\end{equation}
\begin{equation}
L_l = \frac{1}{M} \sum_{k=1}^M \left( y_{\text{pre},l}^{(k)} - y_{\text{true},l}^{(k)} \right)^2
\end{equation}
\begin{equation}
\omega_l =
\begin{cases}
\alpha + (1-\alpha)\dfrac{\Delta L}{L_r + L_i}, & \text{if } L_l > L_{\bar{l}}, \\
1 - \omega_{\bar{l}}, & \text{otherwise}.
\end{cases}
\quad \text{for } l \in \{r, i\}
\end{equation}
where \(M=N(N+1)/2\) represents the number of elements, \(\mathbf{y}_{\text{pre},r}\) and \(\mathbf{y}_{\text{pre},i}\) represent the real and imaginary parts of the predicted values, respectively, while \(\mathbf{y}_{\text{true},r}\) and \(\mathbf{y}_{\text{true},i}\) represents those of the true values. \(\Delta L = |L_r-L_i|\) is the error discrepancy, \(\omega_r\) and \(\omega_i\) are the adaptive weights for the real and imaginary parts, respectively, and \(L_{\text{total}}\) denotes the weighted loss function.

\subsection{Self-Attention Mechanism for Complex-Valued Feature Calibration}
%Inspired by the selective attention mechanism of the primate visual system \cite{b56}, 
To further mitigate the discrepancy between the real and imaginary parts,
a self-attention module is designed in the proposed framework. It enables the network to model their interactions, focus on task-relevant features, and adaptively reweight its contributions. The attention pooling process is described as follows \cite{b54, b55}:
\begin{equation}
f(x)=\sum_{i=1}^n \alpha(x,x_i)y_i
\label{eq}
\end{equation}
where \(x\) denotes the query, \((x_i, y_i)\) is the key-value pair, and \(\alpha(x, x_i)\) is the attention weight assigned to the corresponding value \(y_i\). %For any query, the attention weights of all key-value pairs in the model are a valid probability distribution: they are non-negative, and the sum is 1.
For any given query, the attention weights assigned to all key-value pairs constitute a valid probability distribution, as they are non-negative and sum to one.
\begin{figure}[!t]
\centerline{\includegraphics[width=0.85\columnwidth]{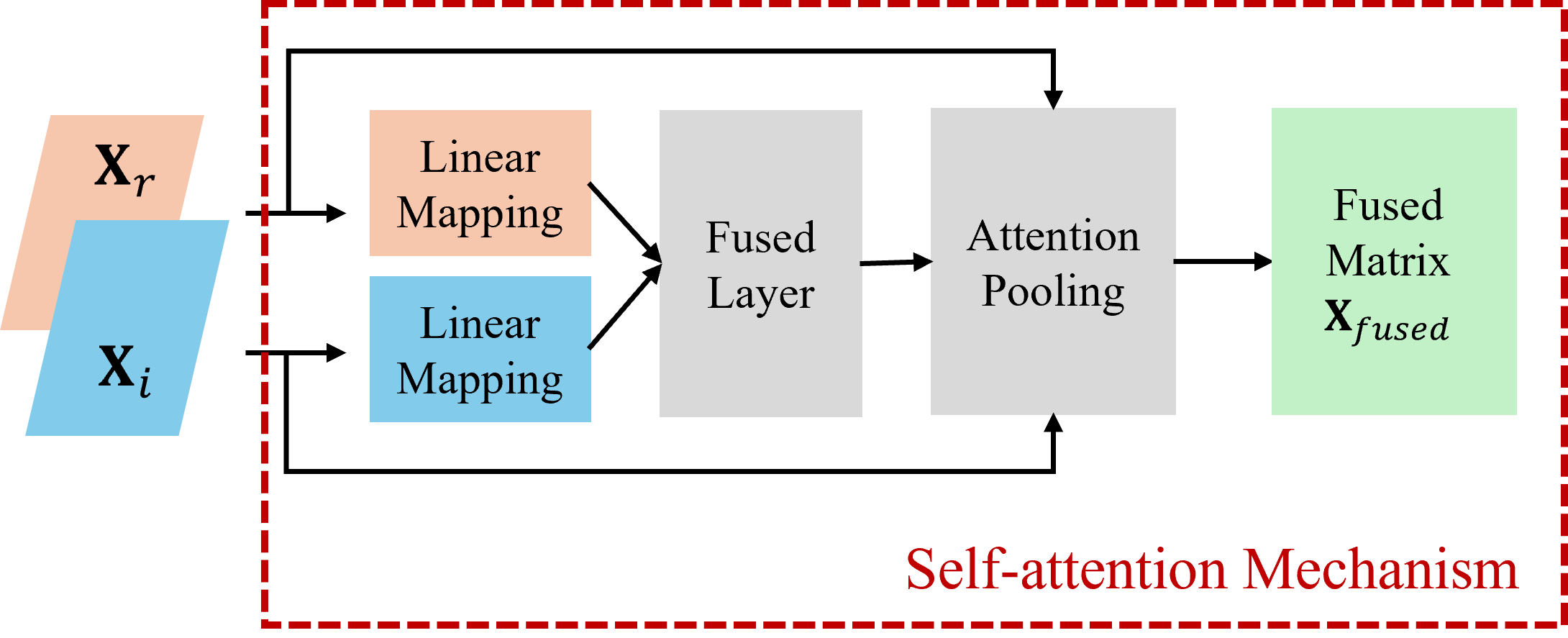}}
\caption{Framework of the proposed self-attention mechanism.}
\label{fig4}
\end{figure}

Fig. \ref{fig4} illustrates the proposed self-attention mechanism with complex-valued inputs. Herein, the real part \(\mathbf{X}_r\) and imaginary part \(\mathbf{X}_i\) of the input are first processed through separate linear mapping layers. These mapped representations are then combined in the fused layer, followed by an attention pooling operation. The final output, \(\mathbf{X}_\textbf{fused}\), is the fused matrix, which integrates both the real and imaginary components using the self-attention mechanism. 

Mathematically, \(\mathbf{X}_r\) and \(\mathbf{X}_i\) are processed by a linear transform via a fully connected layer, which is described as:
\begin{equation}
\mathbf{X}_l'=f(\mathbf{W}_l\mathbf{X}_l+\mathbf{b}_l)
\label{eq}
\quad \text{for } l \in \{r, i\}
\end{equation}
where \(\mathbf{X}_r'\) and \(\mathbf{X}_i'\) respectively refer to features of \(\mathbf{X}_r\) and \(\mathbf{X}_i\), \(\mathbf{W}_r\) and \(\mathbf{W}_i\) are weight matrices of the fully connected layer, and \(f\) is the ReLU function.

Then, \(\mathbf{X}_r'\) and \(\mathbf{X}_i'\) are concatenated together to form a combined feature representation \(\mathbf{X}'=(\mathbf{X}_r', \mathbf{X}_i')\) that contains information about both \(\mathbf{X}_r\) and \(\mathbf{X}_i\). Next, the attention weight \(\mathbf{\alpha}\) is calculated by the SoftMax function:
\begin{equation}
[\mathbf{\alpha_r}, \mathbf{\alpha_i}]=\textbf{SoftMax}(\mathbf{W}_a\cdot\mathbf{X}'+\mathbf{b}_a)
\label{eq}
\end{equation}
where \(\mathbf{\alpha}_r\) and \(\mathbf{\alpha}_i\) represent the attention weights of the real part and imaginary part, respectively, satisfying \(\mathbf{\alpha}_r+\mathbf{\alpha}_i=\mathbf{1}\), \(\mathbf{b}_a\) is the bias, \(\mathbf{W}_a\) is weights, and SoftMax function can be expressed as:
\begin{equation}
\textbf{SoftMax}(\mathbf{z_i})=\frac{e^{z_i}}{\sum_{j=1}^n e^{z_j}}
\label{eq}
\end{equation}
Here \(\mathbf{z} = [z_1, z_2,...,z_n]\) is the input vector, and SoftMax satisfies that the value range of each term is [0,1] and the sum of each term is 1. Herein, the feature-fused matrix \(\mathbf{X}_{\text{fused}}\) can be written as:
\begin{equation}
\mathbf{X}_{\text{fused}} = \mathbf{\alpha}_r\mathbf{X}_r'+\mathbf{\alpha}_i\mathbf{X}_i'
\label{eq29}
\end{equation}
It is clear that attention-based fusion adaptively weights the real and imaginary components, effectively capturing their correlations and enhancing the representation of features.

\subsection{Physics-Aware Convolution Kernel for Feature Extraction}
To extract physically meaningful features from the impedance matrix, we design a physics-aware convolution kernel that integrates domain-specific priors, such as symmetry, central emphasis, and distance-based decay. This design enhances the network’s ability to capture the structured physical patterns that naturally arise from electromagnetic interactions.
Specifically, the convolution kernel can be denoted as \(\mathbf{K}\), with a size of \(3 \times 3\) or larger. The aim is to design a kernel where the weights \(\mathbf{K}(i,j)\) decay as the distance \(d = |i-j|\) from the main diagonal increases, ensuring a spatially adaptive filtering effect. The design principle includes the following two steps.

\textit{Step~1}: \textbf{Design weight decay function}. A common decay function is exponential decay, where the weight has an exponential relationship with the distance, and this function can be described as \(\mathbf{K}(i,j)=
    \begin{cases} 
    \frac{e^{-\alpha|i-j|}}{d(i,j)}, & \text{if } (i, j) \neq (c, c) \\
    1, & \text{if } (i, j) = (c, c)
    \end{cases}\), where \(\alpha > 0\) is the decay factor that controls the speed at which weights decay with distance, \(c\) represents the central index of the matrix, and \(d(i,j)\) represents the Manhattan distance from element \((i,j)\) to center \((c,c)\), which is described as \(d(i,j) = |i-c|+|j-c|\).

\textit{Step~2}: \textbf{Kernel normalization}. To avoid scale distortion during feature extraction, the convolution kernel is normalized such that its elements sum to \(1\), as given by \(\mathbf{K}_{\text{norm}} = \frac{\mathbf{K}}{\sum\mathbf{K}}\).
\begin{comment}
\begin{enumerate}
    \item \textbf{Design weight decay function}.
    A common decay function is exponential decay, where the weight has an exponential relationship with the distance, and this function can be described as:
    \begin{equation}
    \mathbf{K}(i,j)=
    \begin{cases} 
    \frac{e^{-\alpha|i-j|}}{d(i,j)}, & \text{if } (i, j) \neq (c, c), \\
    1, & \text{if } (i, j) = (c, c).
    \end{cases}
    \label{eq}
    \end{equation}
    where \(\alpha > 0\) is the decay factor that controls the speed at which weights decay with distance. \(c\) represents the central index of the matrix, and \(d(i,j)\) represents the Manhattan distance from element \((i,j)\) to center \((c,c)\), which is described as\(d(i,j) = |i-c|+|j-c|\).
    \item \textbf{Kernel normalization}
    To avoid scale distortion during feature extraction, the convolution kernel is normalized such that its elements sum to \(1\), as given by \(\mathbf{K}_{\text{norm}} = \frac{\mathbf{K}}{\sum\mathbf{K}}\).
\end{enumerate}
\end{comment}
By applying this specifically designed convolution kernel, the output of the convolution operation becomes a series of locally weighted sums, effectively capturing the strong coupling patterns embedded in the impedance matrix.
%By using such a designed convolution kernel, the output of the convolution operation will be a series of locally weighted sums, which can reflect the strong coupling patterns in the matrix.

\subsection{LSTM for Modeling Variable-Length Impedance Matrices}
%In MoM, the number of basis functions required for the discretization process depends on the geometrical and electrical characteristics of the analyzed structure, resulting in input sequences of varying lengths. To accommodate this variability, LSTM is leveraged to model and predict the port impedance matrix. The LSTM's inherent ability to handle variable-length sequences \cite{b61} makes it well-suited for this task. To explain how LSTM networks approximate the mapping from the Green's function matrix (\(\mathbf{G}_0\)) to the impedance matrix (\(\mathbf{Z}\)) of MoM and the port impedance matrix \(\mathbf{Z}_{\text{port}}\), it is essential to establish the mathematical connection between the discrete MoM formulation and the sequence modeling mechanism of LSTM.

In the MoM, the number of basis functions required for discretization depends on the geometrical and electrical properties of the structure, resulting in input sequences of varying lengths. To handle this variability, we employ an LSTM network to model and predict the port impedance matrix. The LSTM's ability to process variable-length sequences~\cite{b61} makes it particularly suitable for this task. 
To understand how the LSTM approximates the mapping from the Green's function matrix (\(\mathbf{G}_0\)) to the MoM impedance matrix (\(\mathbf{Z}\)) and the port impedance matrix (\(\mathbf{Z}_{\text{port}}\)), we first establish the mathematical link between the discrete MoM formulation and the sequential modeling capability of the LSTM.

Mathematically, (\ref{eq11}) can be rewritten in discrete form:
\begin{align}
Z_{mn}  = \sum_{m,n} w_{mn} G_{mn}
\label{eq33}
\end{align}
where \(G_{mn}\) denotes the value of the Green's function evaluated at the discrete point \((m, n)\), \(w_{mn}\) represents the corresponding weighting coefficient, which is expressed as:
\begin{align}
w_{mn} & = jkZ_0\Delta \mathbf{r} \Delta\mathbf{r}'\Big(\psi_m(\mathbf{r})\psi_n(\mathbf{r}')-\frac{1}{k^2}\nabla\psi_m(\mathbf{r})\nabla' \cdot\psi_m(\mathbf{r}) \Big)
\label{eq34}
\end{align}
Assume that the antenna array consists of \(M\) dipoles, each divided into into \(N\) segments, resulting in a total of \(T=M\times N\) basis functions. In (\ref{eq13}), \(\mathbf{Z} \in \mathbb{C}^{T \times T}\), \(\mathbf{I} \in \mathbb{C}^T\), and \(\mathbf{V} \in \mathbb{C}^T\). To compute the port impedance matrix, the modified nodal analysis~\cite{b62} method is introduced, and the system equation is expanded as follows:
\begin{equation}
\begin{bmatrix}
\mathbf{Z} & -\mathbf{M}^T \\
\mathbf{M} & \mathbf{0}
\end{bmatrix}
\begin{bmatrix}
\mathbf{I} \\
\mathbf{V}_p
\end{bmatrix}
=
\begin{bmatrix}
\mathbf{0} \\
\mathbf{I}_p
\label{eq35}
\end{bmatrix}
\end{equation}
where \(\mathbf{M} \in \mathbb{R}^{N\times T}\) is the port selection matrix, 
with entries of 1 at the positions corresponding to the selected port segments and 0 elsewhere,
%having a value of 1 at the position of the corresponding port segment, and 0 elsewhere, 
\(\mathbf{V}_p \in \mathbb{C}^N\) denotes the port voltage vector, \(\mathbf{I}_p \in \mathbb{C}^N\) is the port current vector. By solving (\ref{eq35}), we have
\begin{align}
\mathbf{M}\mathbf{Z}^{-1}\mathbf{M}^T\mathbf{V}_p=\mathbf{I}_p
\label{eq}
\end{align}
Accordingly, the port impedance matrix \(\mathbf{Z}_{\text{port}}\) can be defined as:
\begin{align}
\mathbf{Z}_{\text{port}}=\Big(\mathbf{M}\mathbf{Z}^{-1}\mathbf{M}^T\Big)^{-1}
\label{eq}
\end{align}

Then, the processed Green's function matrix \(\mathbf{G}_0\) is reshaped row-wise into a sequence and fed into the LSTM to approximate the port impedance. This transformation can be expressed as $\mathbf{X} = [\mathbf{g}_1,...\mathbf{g}_i,...,\mathbf{g}_N]$, where \(\mathbf{g}_i, i=1,...,N\) represents the \(i\)th row vector of the Green's function \(\mathbf{G}_0\). The cell state update of the LSTM network is generated by
\begin{align}
\mathbf{c}_t & = \mathbf{f}_t \odot \mathbf{c}_{t-1} + \mathbf{i}_t \odot \tilde{\mathbf{c}}_t
\label{eq}
\end{align}
where \(\mathbf{f}_t\) is the forget gate that controls the retention ratio of the historical cell state \(\mathbf{c}_{t-1}\), \(\mathbf{i}_{t}\) is the input gate that regulates the incorporation of the candidate state \(\tilde{c}_t\), the candidate cell state is computed as \(\tilde{c}_t=\)tanh\(\mathbf{W}_c[\mathbf{h}_{t-1}, \mathbf{g}_t] + \mathbf{b}_c\), and \(\odot\) denotes the Hadamard product. The cell state at time \(t\) is recursively defined, and its general form can be derived via mathematical induction:
\begin{align}
\mathbf{c}_t & = \prod_{k=1}^t\mathbf{f}_k \odot \mathbf{c}_{0} + \sum_{\tau = 1}^t\Big( \prod_{k=\tau+1}^t\mathbf{f}_k \Big)\odot \mathbf{i}_{\tau} \odot \tilde{\mathbf{c}}_{\tau}
\label{eq40}
\end{align}

Since \(\mathbf{c}_0=\mathbf{0}\), (\ref{eq40}) simplifies to:
\begin{align}
\mathbf{c}_t & = \sum_{\tau = 1}^t\Big(\prod_{k=\tau+1}^t\mathbf{f}_k\Big)\odot \mathbf{i}_{\tau} \odot \tilde{c}_{\tau}
\label{eq41}
\end{align}
where \(\mathbf{c}_t\) represents dynamically weighted contributions of inputs at each time step. The hidden state of the LSTM is given by:
\begin{align}
\mathbf{h}_t = \mathbf{o}_t \odot \text{tanh}(\mathbf{c}_t)
\label{eq42}
\end{align}
where $\mathbf{o}_t$ is the output gate controlling the exposure of the cell state. Next, the hidden state at the final time step \(t=N\), denoted as \(\mathbf{h}_N\), is mapped to an impedance output \(\mathbf{Z}'\), which is expressed as
\begin{align}
\mathbf{Z}' = \mathbf{W}_z\cdot\mathbf{h}_N + \mathbf{b}_z 
\label{eq43}
\end{align}
By substituting (\ref{eq41}) and (\ref{eq42}) into (\ref{eq43}), one can obtain
\begin{align}
\mathbf{Z}' & = \mathbf{W}_z\cdot \Big( \mathbf{o}_t \odot \text{tanh}(\mathbf{c}_t)\Big) + \mathbf{b}_z \notag \\
 & = \mathbf{W}_z\cdot \Big( \mathbf{o}_t \odot \text{tanh} \Big(\sum_{t = 1}^N \mathbf{\alpha}_{t} \odot \tilde{c}_{t}\Big)\Big) + \mathbf{b}_z \notag \\
 & = \mathbf{W}_z\cdot \Big( \mathbf{o}_t \odot \text{tanh} \Big(\sum_{t = 1}^N \mathbf{\alpha}_{t} \odot \Big(\text{tanh} \mathbf{W}_c\Big[\mathbf{h}_{t-1}, \mathbf{g}_{t} \Big] \notag \\
  & \quad + \mathbf{b}_c\Big)\Big)\Big) + \mathbf{b}_z
\label{eq44}
\end{align}
where \(\mathbf{\alpha}_{t} = \sum_{t = 1}^N \Big(\prod_{k=t+1}^N\mathbf{f}_k\Big)\odot \mathbf{i}_{t}\). Using the first-order approximation,  we can rewrite (\ref{eq44}) as
\begin{align}
\mathbf{Z}' & \approx \mathbf{W}_z\cdot \Big( \mathbf{o}_t \odot \sum_{t = 1}^N \mathbf{\alpha}_{t} \odot \Big(\mathbf{W}_c\Big[\mathbf{h}_{t-1}, \mathbf{g}_{t}\Big] + \mathbf{b}_c\Big)\Big) + \mathbf{b}_z \notag \\
 & =\sum_{t = 1}^N \Big(\mathbf{W}_z\cdot\mathbf{o}_t\odot\mathbf{\alpha}_{t} \odot \Big(\mathbf{W}_c\Big[\mathbf{h}_{t-1}, \mathbf{g}_{t}\Big] + \mathbf{b}_c\Big)\Big) + \mathbf{b}_z \notag \\
 & = \sum_{t = 1}^N \mathbf{\beta}_t\Big[\mathbf{h}_{t-1}, \mathbf{g}_{t}\Big] + \mathbf{b}_z'
\label{eqgreen}
\end{align}
Herein \([\mathbf{h}_{t-1}, \mathbf{g}_{t}]\) is a vector-valued function of the input \(\mathbf{X}\), \(\mathbf{\beta}_t = \mathbf{W}_z\cdot\mathbf{o}_t\odot\mathbf{\alpha}_{t} \odot \mathbf{W}_c\) corresponds to \(w_{mn}\) defined in (\ref{eq34}), and \(\mathbf{b}_z' = \mathbf{b}_z +\sum_{t = 1}^N\mathbf{W}_z\cdot\mathbf{o}_t\odot\mathbf{\alpha}_{t} \odot \mathbf{b}_z\) denotes the bias, enabling the model to more flexibly capture the complex relationship between inputs and outputs. (\ref{eq33}) can be approximated by (\ref{eqgreen}), supporting the assertion that LSTM possesses universal approximation capability and can approximate any non-linear function \cite{b61}. This characteristic renders LSTM particularly suitable for computing the impedance matrix. In a nutshell, the proposed method enables efficient and scalable modeling of variable-length impedance matrices of dipole antenna arrays. %The performance and accuracy of this method are evaluated in the next section.

\section{Numerical examples and discussion}
In this section, several numerical examples, including linear antenna arrays with different numbers of elements and spacings, are used to verify the effectiveness of the proposed method. %All experiments are performed on a workstation with the Intel (R) Core (TM) i7-12700 2.10GHZ.
\subsection{Performance of PANN for Green's Function}
\begin{figure}[!t]
\centerline{\includegraphics[width=0.85\columnwidth]{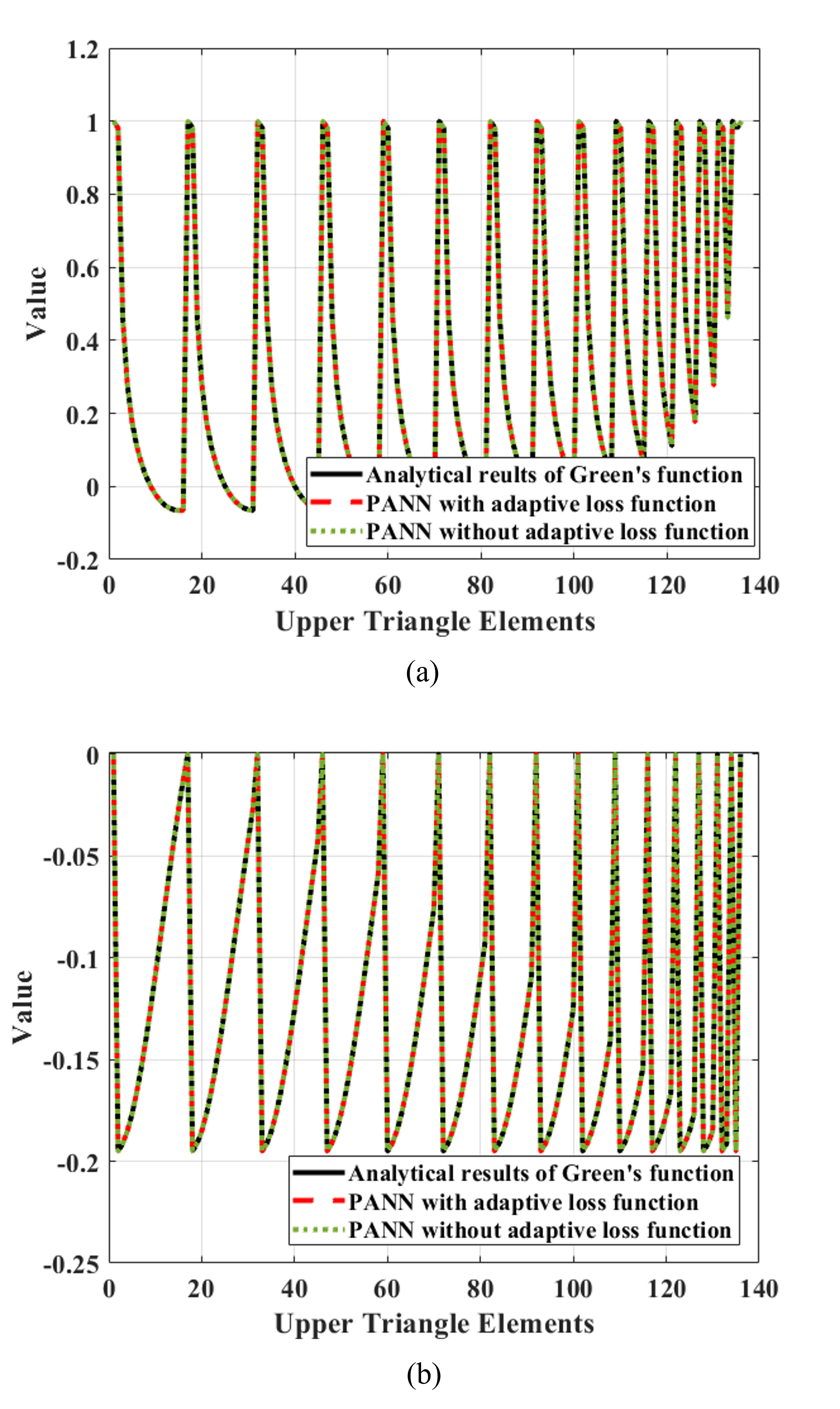}}
\caption{Comparison of real part (a) and imaginary part (b) prediction for Green's function using analytical computation, PANN prediction, and PANN with an adaptive loss function.}
\label{fig5}
\end{figure}

For verification, we first evaluate the performance of PANN in approximating the frequency-independent Green's function.
The number of discretization segments in the MoM analysis is set to 16, and the number of output neurons in PANN is set to \(16\times16=256\). Additionally, based on the symmetric property of the Green's function matrix, it is sufficient to evaluate only the upper triangular part of the Green's function matrix, including the diagonal elements. Fig. \ref{fig5} presents a comparison between the analytical solution and the predictions obtained from PANN, with and without the adaptive loss function, for both the real and imaginary components of the Green's function matrix. The \(x\)-axis corresponds to  the upper triangular elements of the Green's function matrix, while the \(y\)-axis indicates their respective values. It can be observed that both PANN models achieve high accuracy in approximating the analytical results for both real and imaginary parts, demonstrating the effectiveness of the PANN. 

\begin{figure}[!t]
\centerline{\includegraphics[width=0.85\columnwidth]{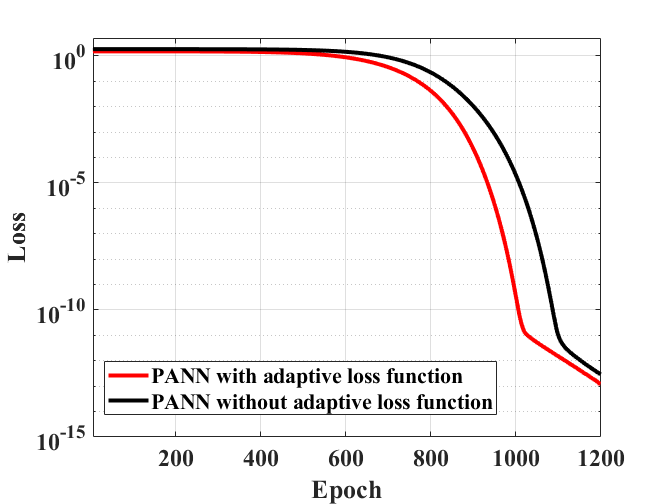}}
\caption{Loss function comparison between PANN and PANN with adaptive loss function.}
\label{fig7}
\end{figure}

For further validation, the loss function curves of two PANN models are compared in Fig. \ref{fig7}. The \(x\)-axis represents the number of training epochs, while \(y\)-axis shows the corresponding loss values on a logarithmic scale. Clearly, the loss functions of both PANN models converge iteratively with low error, indicating the ability of PANN to accurately predict the Green's function matrix in the absence of frequency-dependent terms. It is worth noting that PANN with the adaptive loss function (red line) exhibits a faster loss decay, particularly after 580 epochs, indicating improved convergence and numerical stability. 
Hence, we can conclude that the effectiveness of dynamically balancing errors between the real and imaginary parts in accelerating training while preserving high prediction accuracy for Green's function.

In addition, Table I highlights the exceptional performance of the PANN with the adaptive loss function compared to three other methods in key aspects. Unlike ANN, which requires extensive labeled data through simulation, both PANN and its adaptive variant are unsupervised and fully label-free. Although the adaptive loss introduces slight computational overhead compared to standard PANN, it significantly improves fitting accuracy, achieving an error as low as $10^{-13}$. Besides, once trained, the neural network can rapidly infer accurate solutions with minimal computational cost, eliminating the need for iterative solvers. During inference, all neural network-based methods outperform formula-based computation in speed, with the adaptive PANN variant being the fatest. Overall, the proposed method offers superior accuracy, faster convergence, and eliminates the need for labeled data, making it a robust and efficient approach for fitting the Green's function without frequency-dependent terms. 

\begin{table}[t]
    \centering
    \caption{Performance comparison of different models (ANN, PANN, and PANN combined with adaptive loss function).}
    \label{tab:performance}
    \renewcommand{\arraystretch}{1.2}
    \resizebox{\columnwidth}{!}{ % 缩放到列宽
    \begin{tabular}{c c c c}
        \toprule
        & ANN & PANN & \makecell{PANN + \\ Adaptive Loss Function} \\ 
        \midrule
        Dataset Size  & 1000  & \textbf{-}  & \textbf{-}  \\
        Training iteration  & 5000  & \textbf{1200}  & \textbf{1200}  \\
        Training Time (s)  & 1181  & \textbf{589}  & 620  \\ 
        Inference Time (s)  & 0.00322  & 0.00101  & \textbf{0.00089}  \\ 
        MSE   & 0.2860  & \(5\times10^{-12}\)  & \(\mathbf{10^{-13}}\)  \\ 
        \bottomrule
    \end{tabular}
    }
\end{table}

\subsection{Performance of PC-LSTM on Modeling Impedance Matrix}
\begin{figure}[!t]
\centerline{\includegraphics[width=0.85\columnwidth]{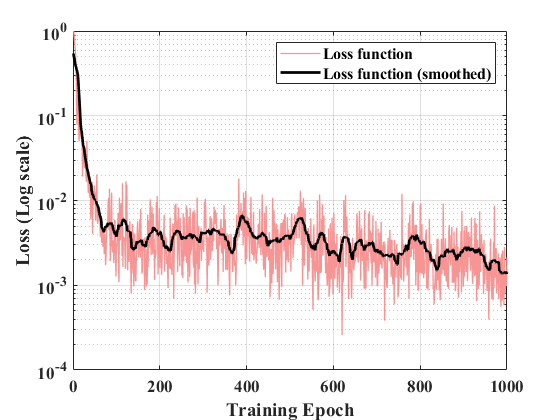}}
\caption{Loss function of PC-LSTM.}
\label{fig8}
\end{figure}

%PC-LSTM is used to construct the port impedance matrix of a two-element linear antenna array, where each element is a half-wavelength dipole antenna.
To evaluate the performance of PC-LSTM in modeling the impedance matrix, we investigate the port impedance matrix of a two-element linear antenna array, where each element is a half-wavelength dipole.
In this work, a dipole antenna array with length \(l = 6.25\) cm, width \(r=0.05\) cm, and an inter-element spacing \(d = 6.25\) cm is selected as a representative test case. The ground truth is generated using the Antenna Toolbox from MATLAB. The trained PANN takes \(d_i\), \(l\), \(r\), and \(f\) as inputs to predict \(\mathbf{G}_r\) and \(\mathbf{G}_i\), each with a dimension of \(32\times 32\). These matrices, along with a fused feature representation \(\mathbf{G}_{\text{fused}}\), are fed into PC-LSTM, yielding an input dimension of \(3072\). The output consists of three neurons corresponding to unique entries of the symmetric port impedance matrix. In addition, this network architecture comprises one convolutional layer followed by four LSTM hidden layers.
Fig.~\ref{fig8} shows both raw and smoothed loss curves, and it can be seen that the error decreases rapidly in early iterations and gradually stabilizes, with the minimum error reaching the order of \(10^{-3}\). This convergence confirms the effectiveness of PC-LSTM in learning the port impedance matrix, with the final prediction error remaining within an acceptable range.

\begin{figure}[!t]
\centerline{\includegraphics[width=0.9\columnwidth]{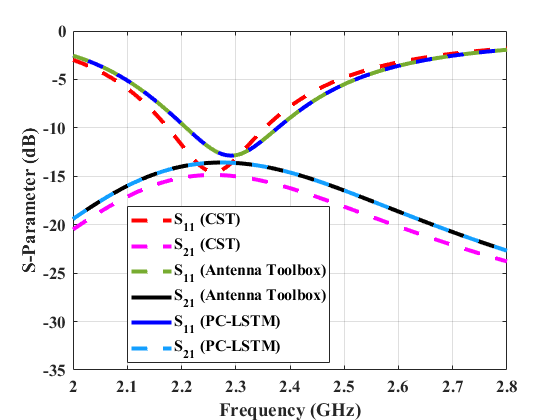}}
\caption{Comparison of S-Parameters using CST, Antenna Toolbox, and PC-LSTM.}
\label{fig9}
\end{figure}

To further validate the effectiveness of the proposed method, the S-parameters, including \(S_{11}\) and \(S_{21}\), predicted by PC-LSTM are compared with those obtained from CST full-wave simulations and MATLAB's Antenna Toolbox, as illustrated in Fig.~\ref{fig9}.
It can be seen that, within the frequency range of 2–2.8 GHz, the S-parameters predicted by PC-LSTM closely match those obtained from the Antenna Toolbox and exhibit trends and resonance points that are highly consistent with CST simulation results, further validating our proposed method. Herein, small discrepancies between the PC-LSTM and CST results are mainly attributed to differences in modeling assumptions. The PC-LSTM model is trained on data from MATLAB's Antenna Toolbox, which assumes ideal dipoles with zero thickness and no feed gap. In contrast, CST explicitly accounts for these physical features, resulting in slight variations in the predicted S-parameters. %These variations in modeling fidelity result in observable differences, as shown in Fig.~\ref{fig9}. 

In addition, we consider two representative cases to further verify the accuracy of the port impedance matrix extracted by the proposed PC-LSTM model. Specifically, cases 1 and 2 correspond to dipole arrays with \(d=0.052\lambda\) and \(d=0.206\lambda\), respectively. Both cases operate at \(3\,\text{GHz}\), with identical dimensions of \(l = 0.5\lambda\) and \(r = 0.002\lambda\), as summarized in Table~II.
For Case~1, the self-impedance \(Z_{11}\) predicted by PC-LSTM shows excellent agreement with the Antenna Toolbox and acceptable deviation from CST, with the relative error of \(0.09\%\) and \(13.1\%\), respectively. As previously discussed, the discrepancy with CST arises primarily from structural differences, remain within an acceptable range for practical applications. In terms of computational efficiency, the proposed method achieves a speedup of over 3x compared to the Antenna Toolbox and more than 7x compared to CST, demonstrating both its accuracy and computational advantage.  
For Case~2, the PC-LSTM prediction of \(Z_{11}\) exhibits a relative error of only \(0.025\%\) compared to the Antenna Toolbox and \(9.46\%\) relative to CST, both within acceptable tolerance. In terms of computational speed, PC-LSTM achieves a speedup of over 3.5x compared to the Antenna Toolbox and more than 7.9x relative to CST, further confirming its predictive accuracy and computational advantage across different array configurations. 
\begin{comment}
\begin{table}[t]
    \centering
    \color{red}
    \caption{Port Impedance Matrix Comparison of PC-LSTM, Antenna Toolbox, and CST for Two Test Cases}
    \label{tab:impedance_combined}
    \renewcommand{\arraystretch}{1.2}
    \resizebox{\columnwidth}{!}{
    \begin{tabular}{c c c c}
        \toprule
          & PC-LSTM & Antenna Toolbox & CST \\
        \midrule
        \multicolumn{4}{c}{\textbf{Case 1:} \(f = 3\,\text{GHz},\ d = 0.052\lambda,\ l = 0.5\lambda,\ r = 0.002\lambda\)} \\
        \midrule
        \(Z_{11} (\Omega)\) & 87.10 + 39.11i & 87.11 + 39.20i & 90.70 + 52.44i \\
        \(Z_{12} (\Omega)\) & 86.37 + 19.44i & 85.42 + 18.69i & 88.65 + 15.95i \\
        \(Z_{21} (\Omega)\) & 86.37 + 19.44i & 85.42 + 18.69i & 88.65 + 15.95i \\
        \(Z_{22} (\Omega)\) & 87.10 + 39.11i & 87.11 + 39.20i & 90.70 + 52.44i \\
        Time (s)            & \textbf{1.414}& 4.602          & 10.392         \\
        \midrule
        \multicolumn{4}{c}{\textbf{Case 2:} \(f = 3\,\text{GHz},\ d = 0.206\lambda,\ l = 0.5\lambda,\ r = 0.002\lambda\)} \\
        \midrule
        \(Z_{11} (\Omega)\) & 80.56 + 41.56i & 80.55 + 41.58i & 80.42 + 50.54i \\
        \(Z_{12} (\Omega)\) & 54.17 - 29.47i & 53.46 - 30.06i & 49.69 - 34.83i \\
        \(Z_{21} (\Omega)\) & 54.17 - 29.47i & 53.46 - 30.06i & 49.69 - 34.83i \\
        \(Z_{22} (\Omega)\) & 80.56 + 41.56i & 80.55 + 41.58i & 80.42 + 50.54i \\
        Time (s)            & \textbf{1.296} & 4.569          & 10.287         \\
        \bottomrule
    \end{tabular}
    }
\end{table}
\end{comment}
\begin{table}[t]
    \centering
    \caption{Port impedance matrix comparison of CST, Antenna Toolbox, and PC-LSTM for two test cases.}
    \label{tab:impedance_combined}
    \renewcommand{\arraystretch}{1.2}
    \resizebox{\columnwidth}{!}{
    \begin{tabular}{c c c c}
        \toprule
          & CST & Antenna Toolbox & PC-LSTM \\
        \midrule
        \multicolumn{4}{c}{\textbf{Case 1:} \(f = 3\,\text{GHz},\ d = 0.052\lambda,\ l = 0.5\lambda,\ r = 0.002\lambda\)} \\
        \midrule
        \(Z_{11} (\Omega)\) & 90.70 + 52.44i & 87.11 + 39.20i & 87.10 + 39.11i \\
        \(Z_{12} (\Omega)\) & 88.65 + 15.95i & 85.42 + 18.69i & 86.37 + 19.44i \\
        \(Z_{21} (\Omega)\) & 88.65 + 15.95i & 85.42 + 18.69i & 86.37 + 19.44i \\
        \(Z_{22} (\Omega)\) & 90.70 + 52.44i & 87.11 + 39.20i & 87.10 + 39.11i \\
        Time (s)            & 10.392         & 4.602          & \textbf{1.414} \\
        Relative Error (\%) & \multicolumn{1}{c}{—} & 0.09\% & 13.1\% \\
        \midrule
        \multicolumn{4}{c}{\textbf{Case 2:} \(f = 3\,\text{GHz},\ d = 0.206\lambda,\ l = 0.5\lambda,\ r = 0.002\lambda\)} \\
        \midrule
        \(Z_{11} (\Omega)\) & 80.42 + 50.54i & 80.55 + 41.58i & 80.56 + 41.56i \\
        \(Z_{12} (\Omega)\) & 49.69 - 34.83i & 53.46 - 30.06i & 54.17 - 29.47i \\
        \(Z_{21} (\Omega)\) & 49.69 - 34.83i & 53.46 - 30.06i & 54.17 - 29.47i \\
        \(Z_{22} (\Omega)\) & 80.42 + 50.54i & 80.55 + 41.58i & 80.56 + 41.56i \\
        Time (s)            & 10.287         & 4.569          & \textbf{1.296} \\
        Relative Error (\%) & \multicolumn{1}{c}{—} & 0.025\% & 9.46\% \\
        \bottomrule
    \end{tabular}
    }
\end{table}
\subsection{Performance of PC-LSTM on Synthesizing the Large-Scale Port Impedance Matrix}
\begin{figure}[!t]
\centerline{\includegraphics[width=0.85\columnwidth]{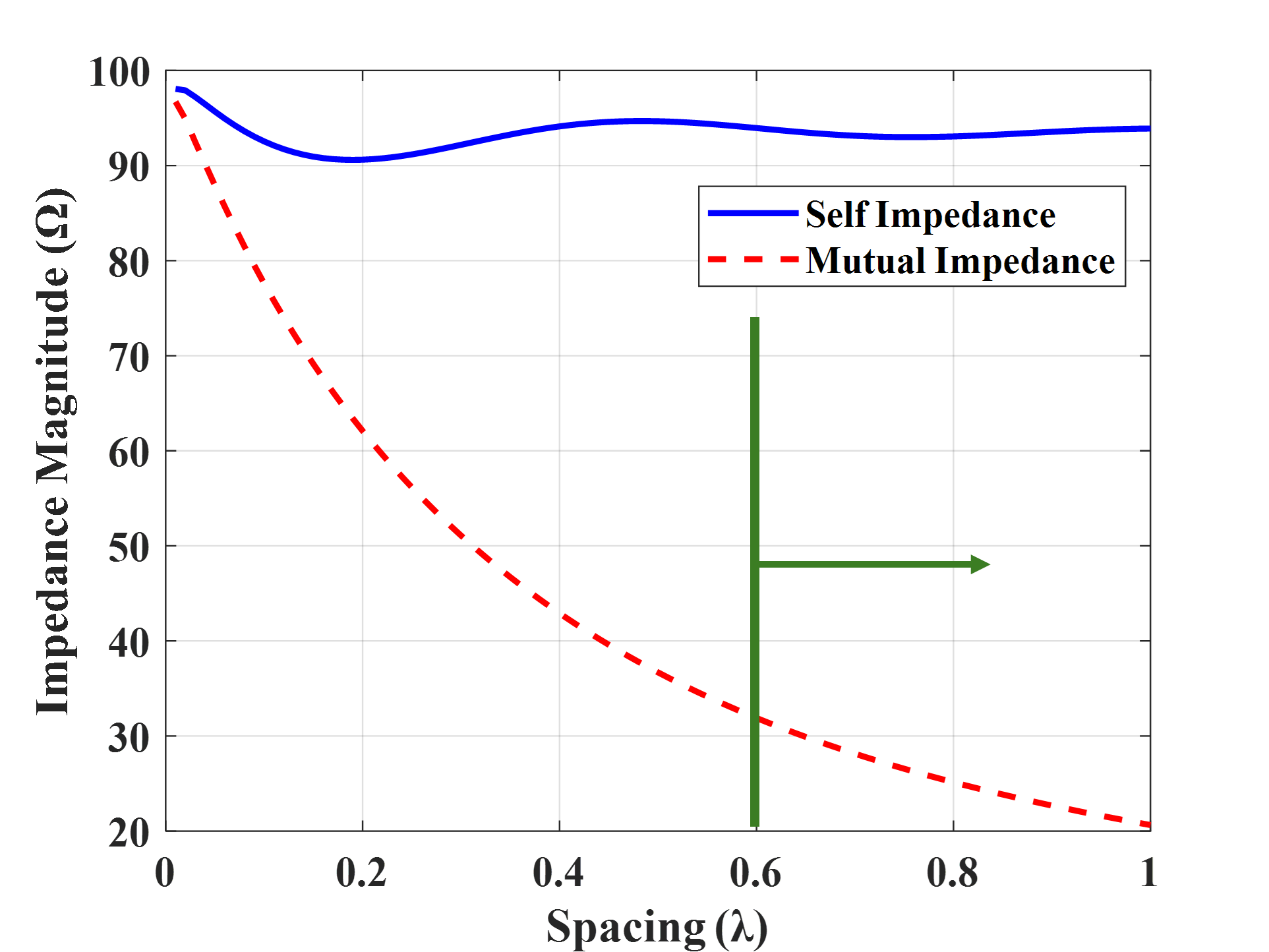}}
\caption{Variations of the self-impedance and mutual impedance variations of a two-element dipole antenna array with length \(l=6.25\) cm and width \(r = 0.05\) cm as the inter-element spacing changes.}
\label{fig10}
\end{figure}

In the case of non-uniform arrays, the spacings are generally non-uniform and some are often smaller than \(0.5\lambda\) \cite{b24}. Based on the port impedance matrix of the two-element dipole antenna array, it can be observed from Fig.~\ref{fig10} that the self-impedance exhibits negligible variation as a function of the inter-element spacing, whereas the mutual impedance demonstrates considerable sensitivity to spacing changes. 
In particular, when the spacing exceeds \(0.6\lambda\), the mutual impedance becomes small enough to be neglected. 

\begin{figure}[!t]
\centerline{\includegraphics[width=1\columnwidth]{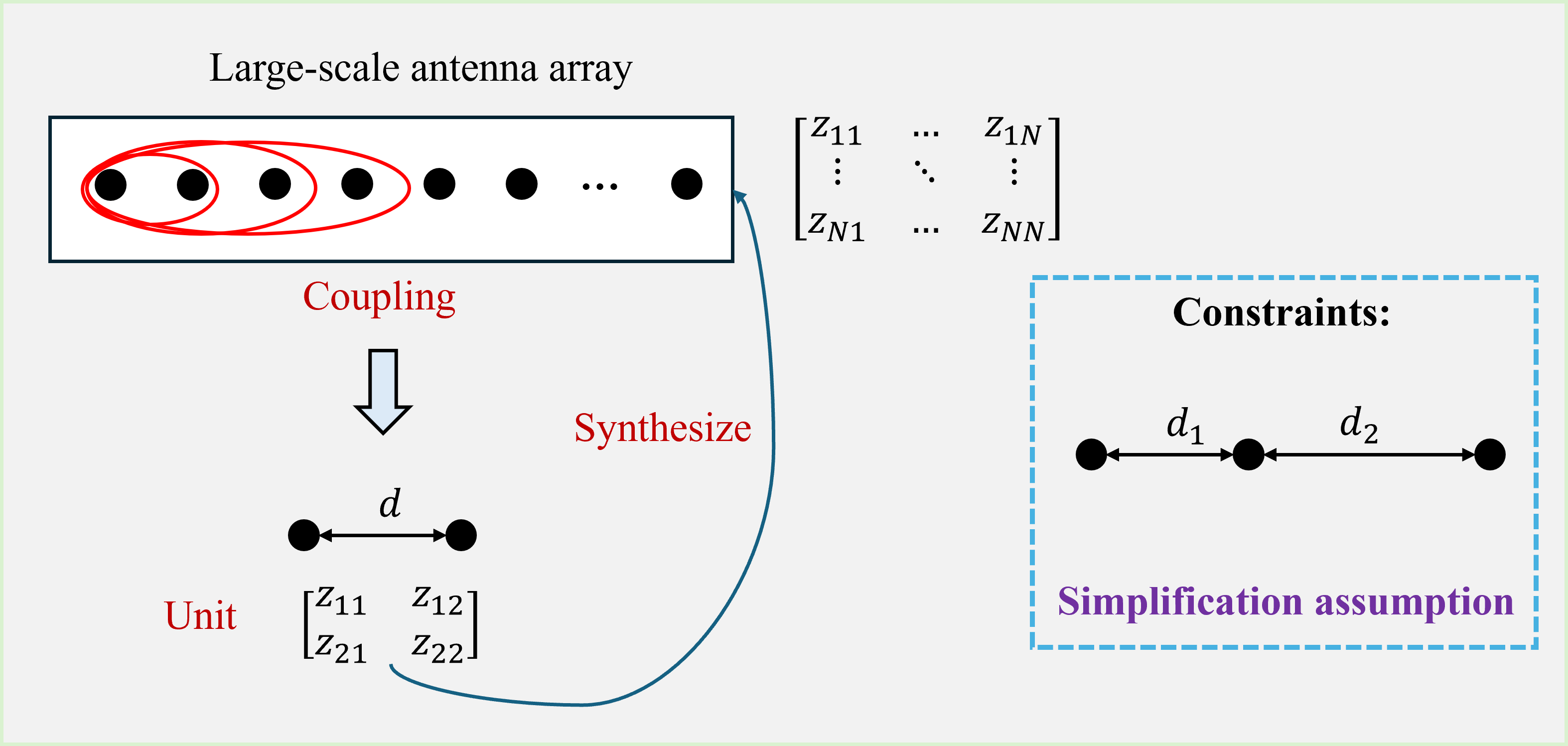}}
\caption{Synthesis of a large-scale antenna array based on a two-element unit model, where coupling is characterized by the impedance matrix and simplification is achieved by imposing constraints on the inter-element spacing.}
\label{fig11}
\end{figure}

As illustrated in Fig.~\ref{fig11}, the antenna array is modeled by a two-element unit, where the coupling behavior is characterized by the corresponding impedance matrix. To simplify the synthesis process, inter-element spacing is subject to the following constraints: (1) \(0.5\lambda \geq d_1 \geq 0.1\lambda\); (2) \(d_1+d_2 \geq 0.6\lambda\). These constraints jointly ensure the validity of the coupling-related assumptions. These constraints effectively reduce the computational complexity of the port impedance matrix for large-scale arrays and align with practical deployment scenarios. 

\begin{figure}[!t]
\centerline{\includegraphics[width=0.85\columnwidth]{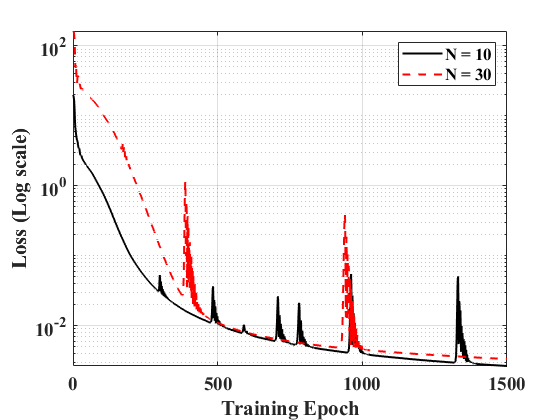}}
\caption{Loss function curves for arrays with 10 elements and (b) 30 elements.}
\label{fig12}
\end{figure}

To validate the proposed PC-LSTM method, we consider examples involving antenna arrays of varying scales.
The geometry of the single-dipole antenna follows the same parameters as described in Section B, Part III, and the ground truth is generated by MATLAB's Antenna Toolbox. Fig.~\ref{fig12} plots the loss curves for arrays with 10 and 30 elements, and it can be seen that the loss curves converge rapidly within a few iterations for both cases, demonstrating the high computational efficiency of PC-LSTM. Also, the training times for the two cases are 10 minutes 10 seconds and 16 minutes 10 seconds, respectively. Once trained, the inference requires only \(1.029\) seconds and \(1.395\) seconds for the two cases, respectively. The minimum loss values achieved are \(1.1\times10^{-3}\) and \(3\times10^{-3}\). Both results fall well within the design tolerance for the synthesis task, thus validating the accuracy of PC-LSTM. These results demonstrate that PC-LSTM can produce solutions both accurately and efficiently.

\begin{figure}[!t]
\centerline{\includegraphics[width=0.85\columnwidth]{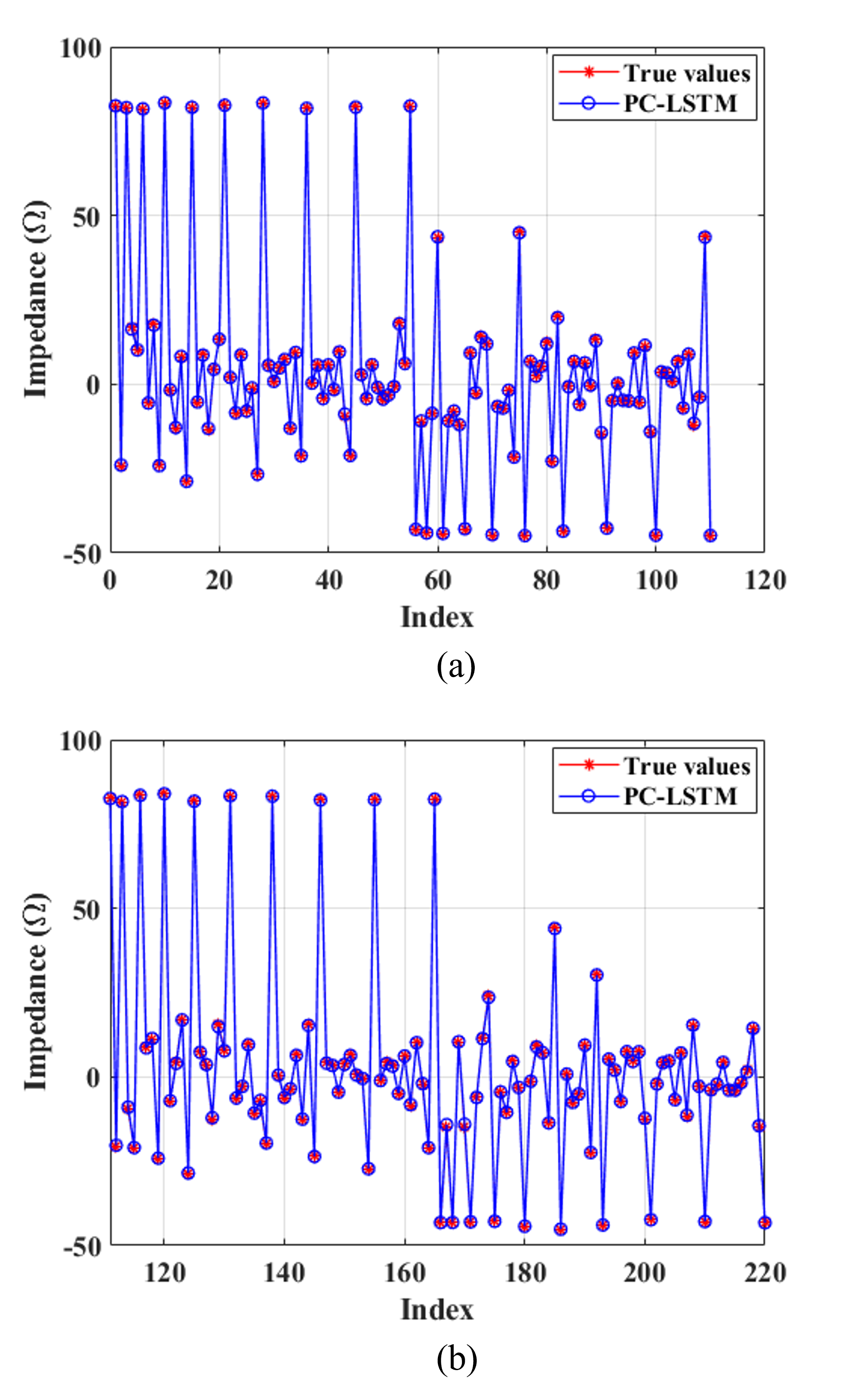}}
\caption{Impedance matrices for two 10-element arrays with different randomly generated inter-element spacings: (a) Case 1 and (b) Case 2.}
\label{fig13}
\end{figure}

The evaluation begins with two small-scale test cases involving 10-element linear dipole arrays, designed to assess the prediction accuracy of the proposed method on arrays with limited aperture and randomly generated inter-element spacings. The predicted port impedance matrices are presented in Fig.~\ref{fig13}. To reduce computation, only the upper triangular part of the symmetric matrix is predicted, with the first 55 indices representing the real part and the last 55 the imaginary part. It can be observed that, regardless of Case 1 and Case 2, the predicted values are consistent with the true values, with negligible error, demonstrating that PC-LSTM can accurately synthesize small-scale arrays.

\begin{figure}[!t]
\centerline{\includegraphics[width=0.85\columnwidth]{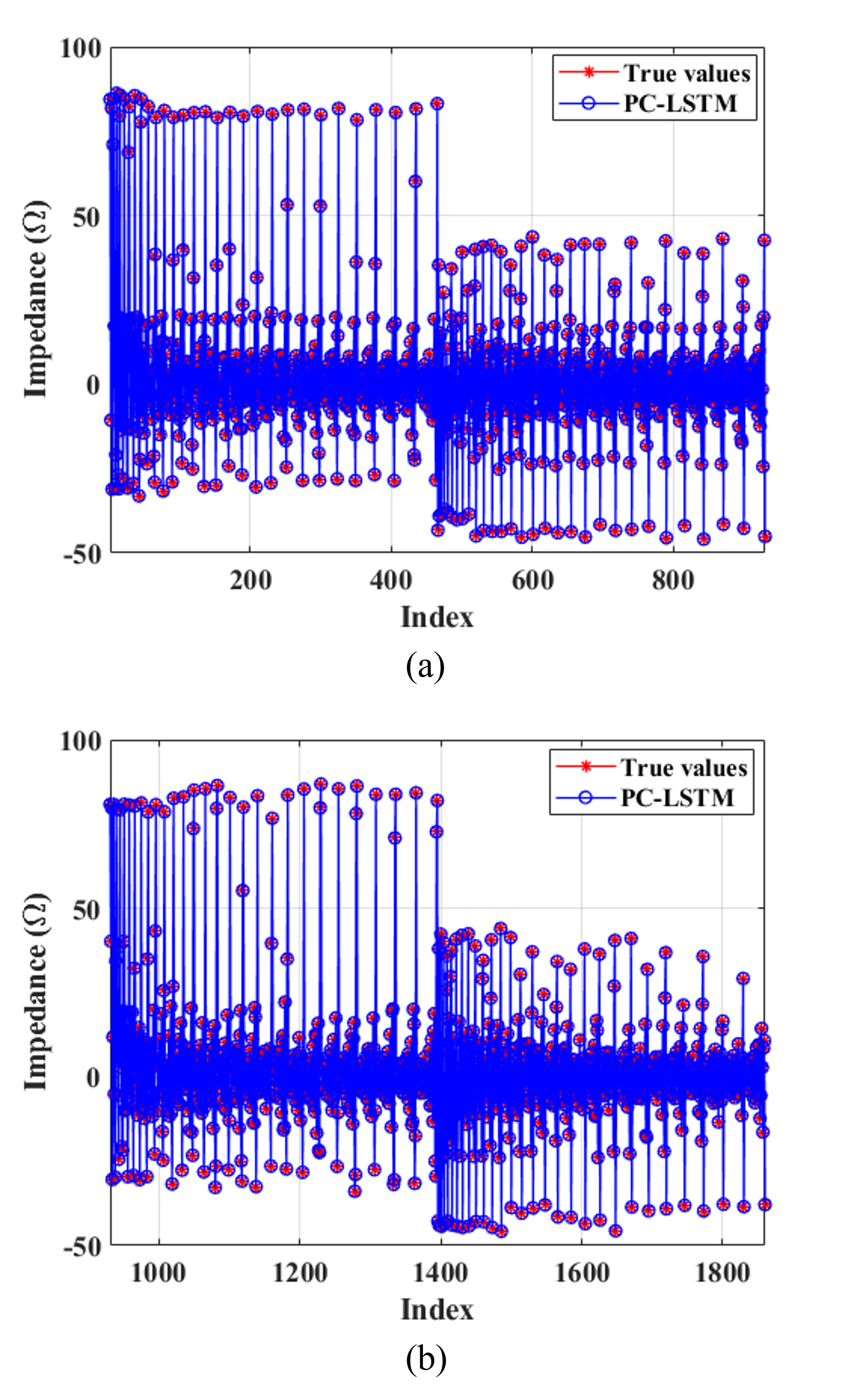}}
\caption{Impedance matrices for two 30-element arrays with different randomly generated inter-element spacings: (a) Case 1 and (b) Case 2.}
\label{fig14}
\end{figure}

Two additional test cases involving 30-element linear dipole arrays with randomly distributed inter-element spacings are considered. The corresponding predicted impedance matrices are shown in Fig.~\ref{fig14}, with the first 465 matrix indices representing the real part and the last 465 matrix indices the imaginary part. 
For both Case 1 and Case 2, the predicted impedance values closely match the ground truth, with negligible discrepancies, demonstrating the effectiveness, robustness, and general applicability of the proposed PC-LSTM method in accurately modeling the impedance characteristics of large-scale linear arrays. 

%As summarized in Table~III, the proposed PC-LSTM method demonstrates superior performance compared to existing approaches (\cite{b63}-\cite{b66}). 

Table~III compares the performance of our proposed method with other similar approaches, including physical preprocessing neural network (PP-NN) \cite{b63}, electromagnetic machine learning (EM-ML) \cite{b64}, masked antoencoders (MAE) \cite{b65}, reinforcement learning deep neural networks (RL-DNN), and PC-LSTM.
Clearly, PC-LSTM achieves high accuracy and physical interpretability with only 100 training samples, significantly reducing data dependency (e.g., 12 times smaller than \cite{b63}). Despite its low model complexity, it maintains high reliability and excellent generalization, outperforming conventional ANN-based methods~\cite{b64,b65} that suffer from limited generalization and high computational costs. This efficiency arises from the integration of physical constraints into the PC-LSTM architecture, enabling robust predictions even in resource-constrained scenarios while maintaining theoretical consistency, which is a critical advantage for real-time electromagnetic systems. 
Therefore, the proposed PC-LSTM method achieves high accuracy, superior generalization, and strong physical interpretability with significantly reduced data requirements, making it highly efficient and reliable for real-time electromagnetic applications.
\begin{table}[t]
    \centering
    \caption{Comparative performance of existing methods and the proposed PC-LSTM across dataset size, physical interpretability, accuracy, model complexity, reliability, and generalization.}
    \label{tab:performance}
    \renewcommand{\arraystretch}{1.2}
    \resizebox{\columnwidth}{!}{ % 缩放到列宽
    \begin{tabular}{lccccc}
        \toprule
        \textbf{Method} & 
        PP-NN~\cite{b63} & 
        EM-ML~\cite{b64} & 
        MAE~\cite{b65} & 
        RL-DNNs~\cite{b66} & 
        PC-LSTM \\
        \midrule
        Dataset Size & 1200 & Not Given & 3104 & 400 & \textbf{100} \\
        Physical Interpretability & \textbf{High} & Medium & Low & Low & \textbf{High} \\
        Accuracy & \textbf{High} & \textbf{High} & Medium & Medium & \textbf{High}\\
        Model Complexity & High & High & High & High & \textbf{Low}\\
        Reliability & Medium & Low & Low & Low & \textbf{High}\\
        Generalization & Low & Low & Low & Low & \textbf{High}\\
        \bottomrule
    \end{tabular}
    }
\end{table}

\section{Conclusion}
To sum up, a physics-aware attention-based machine learning method, namely PC-LSTM, was proposed for efficient modeling of port impedance matrices in dipole antenna arrays. By embedding Green's function into the neural network architecture, it captured mutual coupling effects with minimal reliance on labeled data. In addition, we designed an attention-based scheme to effectively mitigate the discrepancy between the real and
imaginary part components of the port impedance matrix.
Numerical results validated the effectiveness of the proposed PC-LSTM approach, demonstrating its potential for practical applications in antenna array modeling and design.

\bibliographystyle{ieeetr}
\bibliography{references}

\begin{thebibliography}{10}

\bibitem{wen2021compact}
L.~Wen, S.~Gao, B.~Sanz-Izquierdo, C.~Wang, W.~Hu, X.~Ren, and J.~Wu, ``Compact
  and wideband crossed dipole antenna using coupling stub for circular
  polarization,'' {\em IEEE Trans. Antennas Propag.}, vol.~70, no.~1,
  pp.~27--34, 2021.

\bibitem{haupt2010antenna}
R.~L. Haupt, {\em Antenna arrays: a computational approach}.
\newblock John Wiley \& Sons, 2010.

\bibitem{b6}
M.~Li and S.~Cheung, ``A novel calculation-based parasitic decoupling technique
  for increasing isolation in multiple-element {MIMO} antenna arrays,'' {\em
  IEEE Trans. Veh. Technol.}, vol.~70, no.~1, pp.~446--458, 2020.

\bibitem{b13}
C.~A. Alistarh, S.~K. Podilchak, P.~D.~H. Re, T.~M. Str{\"o}ber, Y.~Pailhas,
  C.~Mateo-Segura, M.~Sellathurai, G.~Goussetis, Y.~R. Petillot, J.~S.
  Thompson, {\em et~al.}, ``Sectorized {FMCW MIMO} radar by modular design with
  non-uniform sparse arrays,'' {\em IEEE Journal of Microwaves}, vol.~2, no.~3,
  pp.~442--460, 2022.

\bibitem{b14}
J.~Ding, Z.~Wang, W.~Ma, X.~Wu, and M.~Wang, ``{Tdm}-{mimo} automotive radar
  point-cloud detection based on the 2-d hybrid sparse antenna array,'' {\em
  IEEE Trans. Geosci. Remote Sens.}, vol.~60, pp.~1--15, 2022.

\bibitem{b11}
A.~Kedar, {\em Sparse Phased Array Antennas: Theory and Applications}.
\newblock Artech House, 2022.

\bibitem{b12}
M.~G. Amin, {\em Sparse Arrays for Radar, Sonar, and Communications}.
\newblock John Wiley \& Sons, 2024.

\bibitem{b15}
Q.~Yuan, Q.~Chen, and K.~Sawaya, ``Performance of adaptive array antenna with
  arbitrary geometry in the presence of mutual coupling,'' {\em IEEE Trans.
  Antennas Propag.}, vol.~54, no.~7, pp.~1991--1996, 2006.

\bibitem{b21}
C.~Bencivenni, M.~Ivashina, R.~Maaskant, and J.~Wettergren, ``Design of
  maximally sparse antenna arrays in the presence of mutual coupling,'' {\em
  IEEE Antennas Wirel. Propag. Lett.}, vol.~14, pp.~159--162, 2014.

\bibitem{PreviousTAP}
C.~Wang, Y.~Zhang, S.~Gao, and W.~Liu, ``Design of sparse antenna array using
  physics-aware generative adversarial network,'' {\em IEEE Trans. Antennas
  Propag.}, pp.~1--1, 2025.

\bibitem{10738292}
Y.~Zhang, W.~Xu, A.-L. Jin, M.~Li, P.~Ma, L.~Jiang, and S.~Gao,
  ``Coupling-informed data-driven scheme for joint angle and frequency
  estimation in uniform linear array with mutual coupling present,'' {\em IEEE
  Trans. Antennas Propag.}, vol.~72, no.~12, pp.~9117--9128, 2024.

\bibitem{10858644}
M.~Li, Y.~He, C.~Zhou, Y.~Zhang, and D.~Wu, ``Design of decoupling and pattern
  shaping surface for {MIMO} antennas using the multiport optimization
  method,'' {\em IEEE Trans. Antennas Propag.}, vol.~73, no.~5, pp.~2927--2939,
  2025.

\bibitem{9411296}
M.~Alibakhshikenari, B.~S. Virdee, A.~A. Althuwayb, F.~Falcone, and E.~Limiti,
  ``Interaction suppression technique for high-density antenna arrays for
  mm-wave {5G} {MIMO} systems,'' in {\em 2021 15th European Conference on
  Antennas and Propagation (EuCAP)}, pp.~1--5, 2021.

\bibitem{lu2024generalized}
W.-J. Lu, ``Generalized odd-even mode theory and mode synthesis antenna design
  approach,'' {\em Electromagnetic Science}, vol.~2, no.~1, pp.~1--18, 2024.

\bibitem{zhu2023multimode}
L.~Zhu and N.~Liu, ``Multimode resonator technique in antennas: A review,''
  {\em Electromagnetic Science}, vol.~1, no.~1, pp.~1--17, 2023.

\bibitem{9714820}
D.~Sarkar and Y.~M.~M. Antar, ``{FDTD} computation of space/time integrated
  electromagnetic lagrangian: New insights into design of mutually coupled
  antennas,'' {\em IEEE J. Multiscale Multiphysics Comput. Tech.}, vol.~7,
  pp.~16--22, 2022.

\bibitem{1406246}
J.~Rubio, M.~Gonzalez, and J.~Zapata, ``Generalized-scattering-matrix analysis
  of a class of finite arrays of coupled antennas by using 3-{D} {FEM} and
  spherical mode expansion,'' {\em IEEE Trans. Antennas Propag.}, vol.~53,
  no.~3, pp.~1133--1144, 2005.

\bibitem{b25}
I.~Gupta and A.~Ksienski, ``Effect of mutual coupling on the performance of
  adaptive arrays,'' {\em IEEE Trans. Antennas Propag.}, vol.~31, no.~5,
  pp.~785--791, 1983.

\bibitem{704828}
H.~Yuan, K.~Hirasawa, and Y.~Zhang, ``The mutual coupling and diffraction
  effects on the performance of a {CMA} adaptive array,'' {\em IEEE Trans. Veh.
  Technol.}, vol.~47, no.~3, pp.~728--736, 1998.

\bibitem{b28}
R.~Li, D.~Li, J.~Ma, Y.~Wu, Z.~Gu, L.~Zhang, H.~Ma, H.~Chen, and E.-P. Li,
  ``Adaptive mutual coupling compensation based on efficient characterization
  of coupled antenna arrays,'' {\em IEEE Trans. Electromagn. Compat.}, 2023.

\bibitem{b36}
D.~F. Kelley and W.~L. Stutzman, ``Array antenna pattern modeling methods that
  include mutual coupling effects,'' {\em IEEE Trans. Antennas Propag.},
  vol.~41, no.~12, pp.~1625--1632, 1993.

\bibitem{b37}
A.~R. Barron, ``Universal approximation bounds for superpositions of a
  sigmoidal function,'' {\em IEEE Trans. Inf. Theory}, vol.~39, no.~3,
  pp.~930--945, 1993.

\bibitem{b44}
T.~Liu, B.~Song, F.~Meng, W.~Yang, J.~Li, J.~You, and W.~Lu, ``Rapid estimation
  method for coupling degree of airborne antenna based on quantum neural
  network,'' {\em IEEE Antennas Wirel. Propag. Lett.}, 2024.

\bibitem{b45}
Z.~Wei, Z.~Zhou, P.~Wang, J.~Ren, Y.~Yin, G.~F. Pedersen, and M.~Shen, ``Fully
  automated design method based on reinforcement learning and surrogate
  modeling for antenna array decoupling,'' {\em IEEE Trans. Antennas Propag.},
  vol.~71, no.~1, pp.~660--671, 2022.

\bibitem{b46}
H.~Huang, X.-S. Yang, and B.-Z. Wang, ``Machine-learning-based generative
  optimization method and its application to an antenna decoupling design,''
  {\em IEEE Trans. Antennas Propag.}, vol.~71, no.~7, pp.~6243--6248, 2023.

\bibitem{b42}
J.~Zheng, Q.~Lan, X.~Zhang, W.~Kainz, and J.~Chen, ``Prediction of {MRI} {RF}
  exposure for implantable plate devices using artificial neural network,''
  {\em IEEE Trans. Electromagn. Compat.}, vol.~62, no.~3, pp.~673--681, 2019.

\bibitem{b38}
D.~Li, Y.~Gu, H.~Ma, Y.~Li, L.~Zhang, R.~Li, R.~Hao, and E.-P. Li, ``Deep
  learning inverse analysis of higher order modes in monocone {TEM} cell,''
  {\em IEEE Trans. Microw. Theory Tech.}, vol.~70, no.~12, pp.~5332--5339,
  2022.

\bibitem{b43}
M.~Raissi, P.~Perdikaris, and G.~E. Karniadakis, ``Physics-informed neural
  networks: A deep learning framework for solving forward and inverse problems
  involving nonlinear partial differential equations,'' {\em J. Comput. Phys.},
  vol.~378, pp.~686--707, 2019.

\bibitem{cst2025}
``{CST Studio Suite}.'' \url{https://www.cst.com}.
\newblock Available.

\bibitem{6236032}
Y.~P. Chen, W.~C. Chew, and L.~Jiang, ``A new {G}reen's function formulation
  for modeling homogeneous objects in layered medium,'' {\em IEEE Trans.
  Antennas Propag.}, vol.~60, no.~10, pp.~4766--4776, 2012.

\bibitem{b30}
J.-M. Jin, {\em Theory and computation of electromagnetic fields}.
\newblock John Wiley \& Sons, 2015.

\bibitem{b49}
S.~Rao, D.~Wilton, and A.~Glisson, ``Electromagnetic scattering by surfaces of
  arbitrary shape,'' {\em IEEE Trans. Antennas Propag.}, vol.~30, no.~3,
  pp.~409--418, 1982.

\bibitem{b50}
J.~M. Rius, E.~Ubeda, and J.~Parr{\'o}n, ``On the testing of the magnetic field
  integral equation with {RWG} basis functions in method of moments,'' {\em
  IEEE Trans. Antennas Propag.}, vol.~49, no.~11, pp.~1550--1553, 2001.

\bibitem{b51}
E.~A. Soliman, M.~H. Bakr, and N.~K. Nikolova, ``Neural networks-method of
  moments ({NN-MoM}) for the efficient filling of the coupling matrix,'' {\em
  IEEE Trans. Antennas Propag.}, vol.~52, no.~6, pp.~1521--1529, 2004.

\bibitem{10701557}
Y.~Ping, Y.~Zhang, and L.~Jiang, ``Uncertainty quantification in {PEEC} method:
  A physics-informed neural networks-based polynomial chaos expansion,'' {\em
  IEEE Trans. Electromagn. Compat.}, vol.~66, no.~6, pp.~2095--2101, 2024.

\bibitem{su2025multi}
J.~L. Su, Z.~X. Cai, Y.~Mao, L.~Chen, X.~Y. Yu, Z.~C. Yu, Q.~Ma, S.~Q. Huang,
  J.~Zhang, J.~W. You, {\em et~al.}, ``Multi-dimensional multiplexed
  metasurface for multifunctional near-field modulation by physics-driven
  intelligent design,'' {\em Adv. Sci.}, p.~2503899, 2025.

\bibitem{b53}
H.~B. Barlow, ``Unsupervised learning,'' {\em Neural Comput.}, vol.~1, no.~3,
  pp.~295--311, 1989.

\bibitem{b54}
E.~A. Nadaraya, ``On estimating regression,'' {\em Theory of Probability \& Its
  Applications}, vol.~9, no.~1, pp.~141--142, 1964.

\bibitem{b55}
G.~S. Watson, ``Smooth regression analysis,'' {\em Sankhy{\=a}: The Indian
  Journal of Statistics, Series A}, pp.~359--372, 1964.

\bibitem{b61}
S.~Hochreiter and J.~Schmidhuber, ``Long short-term memory,'' {\em Neural
  Comput.}, vol.~9, no.~8, pp.~1735--1780, 1997.

\bibitem{b62}
C.-W. Ho, A.~Ruehli, and P.~Brennan, ``The modified nodal approach to network
  analysis,'' {\em IEEE Trans. Circuits Syst.}, vol.~22, no.~6, pp.~504--509,
  1975.

\bibitem{b24}
C.~A. Balanis, {\em Antenna theory: analysis and design}.
\newblock John wiley \& sons, 2015.

\bibitem{b63}
Y.~Jiang, S.~S. Yuan, and E.~Wei, ``A novel mutual coupling ann model for mimo
  antennas with physical preprocessing,'' {\em IEEE Antennas Wirel. Propag.
  Lett.}, 2024.

\bibitem{b64}
A.~M. Alzahed, S.~M. Mikki, and Y.~M. Antar, ``Nonlinear mutual coupling
  compensation operator design using a novel electromagnetic machine learning
  paradigm,'' {\em IEEE Antennas Wirel. Propag. Lett.}, vol.~18, no.~5,
  pp.~861--865, 2019.

\bibitem{b65}
H.~Huang, X.-S. Yang, and B.-Z. Wang, ``Machine-learning-based generative
  optimization method and its application to an antenna decoupling design,''
  {\em IEEE Trans. Antennas Propag.}, vol.~71, no.~7, pp.~6243--6248, 2023.

\bibitem{b66}
Z.~Wei, Z.~Zhou, P.~Wang, J.~Ren, Y.~Yin, G.~F. Pedersen, and M.~Shen, ``Fully
  automated design method based on reinforcement learning and surrogate
  modeling for antenna array decoupling,'' {\em IEEE Trans. Antennas Propag.},
  vol.~71, no.~1, pp.~660--671, 2022.

\end{thebibliography}
\end{document}